\documentclass[aps,prx,reprint,superscriptaddress]{revtex4-1} 

\usepackage{graphicx}
\usepackage{hyperref}

\begin{document}


\title{A Unified Approach to Enhanced Sampling}


\author{Michele Invernizzi}
\email{michele.invernizzi@phys.chem.ethz.ch}
\affiliation{Department of Physics, ETH Zurich, c/o Universit\`{a} della Svizzera italiana, Via Giuseppe Buffi 13, 6900 Lugano, Switzerland}
\affiliation{Facolt\`{a} di Informatica, Institute of Computational Science,
National Center for Computational Design and Discovery of Novel Materials (MARVEL), Universit\`{a} della Svizzera italiana, Via Giuseppe Buffi 13, 6900 Lugano, Switzerland}
\author{Pablo M. Piaggi}
\affiliation{Department of Chemistry, Princeton University, Princeton, New Jersey 08540, USA}
\author{Michele Parrinello}
\email{parrinello@phys.chem.ethz.ch}
\affiliation{Department of Chemistry and Applied Biosciences, ETH Zurich, c/o Universit\`{a} della Svizzera italiana, Via Giuseppe Buffi 13, 6900 Lugano, Switzerland,
and Italian Institute of Technology, Via Morego 30, 16163 Genova, Italy}
\affiliation{Facolt\`{a} di Informatica, Institute of Computational Science,
National Center for Computational Design and Discovery of Novel Materials (MARVEL), Universit\`{a} della Svizzera italiana, Via Giuseppe Buffi 13, 6900 Lugano, Switzerland}


\date{\today}

\begin{abstract}
The sampling problem lies at the heart of atomistic simulations and over the years many different enhanced sampling methods have been suggested towards its solution.
These methods are often grouped into two broad families.
On the one hand methods such as umbrella sampling and metadynamics that build a bias potential based on few order parameters or collective variables.
On the other hand, tempering methods such as replica exchange that combine different thermodynamic ensembles in one single expanded ensemble.
We instead adopt a unifying perspective, focusing on the target probability distribution sampled by the different methods.
This allows us to introduce a new class of collective-variables-based bias potentials that can be used to sample any of the expanded ensembles normally sampled via replica exchange.
We also provide a practical implementation, by properly adapting the iterative scheme of the recently developed on-the-fly probability enhanced sampling method [Invernizzi and Parrinello, J. Phys. Chem. Lett. 11.7 (2020)], which was originally introduced for metadynamics-like sampling.
The resulting method is very general and can be used to achieve different types of enhanced sampling.
It is also reliable and simple to use, since it presents only few and robust external parameters and has a straightforward reweighting scheme.
Furthermore, it can be used with any number of parallel replicas.
We show the versatility of our approach with applications to multicanonical and multithermal-multibaric simulations, thermodynamic integration, umbrella sampling, and combinations thereof.
\end{abstract}


\maketitle

\section{Introduction}
Sampling is one of the main challenges in atomistic simulations.
In fact even the most accurate models cannot produce high quality results if the phase space is not properly sampled.
The sampling issue is due to the large gap between the physical macroscopic timescales and the actual time that can be explored in standard atomistic simulations.
This results in an ergodicity problem that can be encountered in fields as varied as materials science, chemistry, and biology.
One facet of this problem is the existence of different metastable states separated by kinetic bottlenecks, that make the transition from one state to another a rare event.
Enhanced sampling methods are a possible solution to this problem.
Instead of extracting configurations from the relevant physical ensemble, these methods create an ad hoc modified ensemble in which the probability of sampling rare events is greatly enhanced.
One kind of such target ensembles is obtained by combining multiple sub-ensembles that differ only for the temperature or some other quantity, a typical example being parallel tempering\cite{Earl2005}.
We shall refer to these ensembles as expanded ensembles\cite{Lyubartsev1992}. 

In the present paper we formulate the problem of generating such expanded ensembles in a way that allows us to use collective-variables-based methods.
We find that the recently developed on-the-fly probability enhanced sampling\cite{Invernizzi2020} (OPES), can be adapted to the scope and provides and efficient implementation.
This method was introduced as an evolution of metadynamics\cite{Laio2002}, since it can provide the same type of enhanced sampling, but presents in most cases a faster convergence and has only few and robust adjustable parameters.
These properties of OPES are retained when it is applied to sample expanded ensembles.
This provides us with a general and reliable method, that can be easily applied to sample many different ensembles.

We accompany this paper with a number of general considerations (Secs.~\ref{S:unified} and \ref{S:optimal}), but the reader mostly interested in the method itself and its practical implementation can go directly to Sec.~\ref{S:targeting}.
Section \ref{S:opes} briefly recalls OPES in its original formulation for metadynamics-like sampling.
We also present a variety of simulations to show the versatility of the new scheme, in particular (Sec.~\ref{S:linearly}) multicanonical, multithermal-multibaric, thermodynamic integration, but also (Sec.~\ref{S:multiumbrellas}) enhanced sampling based on an order parameter, both alone and in combination with the previous ensembles.

\section{A Unified Approach}\label{S:unified}
The most popular approaches to enhanced sampling follow mainly two strategies.
A first one was proposed in a pioneering work by Torrie and Valleau and referred to as umbrella sampling\cite{Torrie1974,Torrie1977}.
This method starts by identifying one or few order parameters, or collective variables (CVs), $\mathbf{s}=\mathbf{s}(\mathbf{x})$, that are function of the microscopic configuration and encode some of the slow modes of the system.
Then a bias potential that is function of the CVs is added to the energy of the system, so that the sampling of the slow modes encoded by the CVs is accelerated.
Many have followed this approach, and nowadays one of the most popular methods in this class is metadynamics\cite{Laio2002}.

A different perspective to enhanced sampling is that of parallel or simulated tempering\cite{Swendsen1986,Marinari1992}.
In this case the idea is to combine in the same generalized ensemble the configurations explored by the system at different temperatures.
This can improve the sampling because at higher temperatures the exploration of the phase space is often more efficient, and the system is less likely to remain stuck in metastable states.
Over the years this approach has been extended and implemented in a variety of different methods, among which replica exchange\cite{Sugita1999} is probably the most widely employed.

These two families of enhanced sampling methods often have been seen as distinct and complementary.
Although there are some papers in which the two perspectives are combined\cite{Bussi2006,Piana2007,Bonomi2010,Nava2015}, typically they have been perceived as hybrid approaches\cite{Abrams2014,Yang2019}.
Here we want to take a closer look at these two families and show that it is possible to provide a unified perspective to the enhanced sampling problem.

For a start we must specify that we are not interested in looking at the specific computational technique the various enhanced sampling methods use, since according to this criterion there would be many more than two families.
There are methods that use a bias potential and others that use specific Monte Carlo moves\cite{Sugita1999}, but also methods that introduce a fictitious dynamics\cite{Rosso2002}, or that focus on directly modifying the atomic forces\cite{Darve2001}, to name just a few.
This kind of classification is of course perfectly legitimate, but we find it of limited relevance for our purposes.

The distinction between the two families cannot be based on the fact that one uses system-specific CVs, while the other makes use of general thermodynamic properties.
For instance it is known that Hamiltonian replica exchange can be used to enhance the fluctuations of any chosen CV\cite{Sugita2000}, and on the other hand that it is possible with metadynamics to use the potential energy itself as CV and sample a multithermal ensemble\cite{Micheletti2004}.

Thus we prefer to focus on the target distribution $p^{tg}(\mathbf{x})$ that the different methods sample.
In fact each enhanced sampling method explicitly or implicitly aims at sampling a specific probability distribution in the configuration space that is not the physical one, but assigns a higher probability to some rare event.
Designing such target distributions so that they are effective is far from trivial, and we can relate the two families of methods to the type of target distribution they imply.

A first class of enhanced sampling methods defines the target distribution by setting a constraint on its marginal distribution along some chosen CVs, $p^{tg}(\mathbf{s})=\int \delta (s(\mathbf{x})-s)\,  p^{tg}(\mathbf{x})\, d\mathbf{x}$.
The most common choice is to impose a uniform marginal, $p^{tg}(\mathbf{s})=const$.
Among the methods that adopt this strategy are adaptive umbrella sampling\cite{Mezei1987} and  metadynamics in its original formulation\cite{Laio2002}. 
Also the Wang-Landau algorithm\cite{Wang2001} and various multicanonical algorithms\cite{Berg1991,Lee1993} chose to sample a flat marginal distribution, using the potential energy as CV.
An interesting case is the one of well-tempered metadynamics\cite{Barducci2008} that aims at sampling an $\mathbf{s}$ distribution that is a smoothed version of the original one.
Contrary to the uniform case and in general to the fixed target case\cite{White2015}, the well-tempered target explicitly depends on the unbiased probability, and is thus not completely known beforehand.
Other kinds of targets are also used in the $1/k$ ensemble\cite{Hesselbo1995} and in nested sampling\cite{Partay2010}.

Another class of methods will be the main focus of this paper and it is the one that aims at sampling the so-called expanded ensembles\cite{Lyubartsev1992}.
These targets are not defined explicitly as a function of some CVs, but rather consist in the sum of overlapping probability distributions.
A typical enhanced sampling technique that targets expanded distributions is for example replica exchange\cite{Sugita1999}.
Expanded ensembles can be obtained by combining the same system at different temperatures, or more in general different Hamiltonians\cite{Sugita2000,Liu2005,Bussi2014}.
They can be sampled also with single replica approaches, such as simulated tempering\cite{Marinari1992}, and integrated tempering\cite{Gao2008,Nymeyer2010}.
Broadly speaking, one could consider as part of this expanded ensemble class also methods like multiple windows umbrella sampling\cite{Kastner2011}, or thermodynamic integration\cite{Vega2008}, where multiple separated ensembles are simulated and then combined into one via some post-processing procedure such as WHAM\cite{Kumar1992}.

It is important to notice that by classifying enhanced sampling methods with respect to $p^{tg}(\mathbf{x})$ we are not implying that methods in the same class are equivalent.
Different methods in fact can use very different strategies to reach their target, each having its own strengths and weaknesses.
However, this target-distribution perspective suggest that there is not a fundamental difference between the two traditional families, and that a unified approach is possible.

From this point of view, a special place is occupied by variationally enhanced sampling (VES)\cite{Valsson2014} and on-the-fly probability enhanced sampling (OPES)\cite{Invernizzi2020}, since in these methods one has to explicitly choose a target distribution.
This makes them particularly suited for developing a unified approach, since they are in principle capable of sampling the targets of both of the two families of enhanced sampling, and also combine them in new ways.
In VES the usefulness of various target distributions has already been explored\cite{Shaffer2016,Invernizzi2017,Debnath2019}.
In particular, a target distribution has been proposed for sampling multithermal-multibaric ensembles\cite{Piaggi2019} and also for combining them with a CV that drives a phase transition\cite{Piaggi2019cv}. 
It is this paper that inspired us to try a generalized unified approach.

Our goal here is to introduce explicitly the expanded ensemble target distribution and show that it can be sampled by using a CV-based bias potential method such as VES or OPES.
In doing so we will introduce the concept of expansion CVs, that allows us to define both the target expanded distribution and the bias needed to sample it.
The method we propose is thus capable of sampling both kind of target distributions, those typical of replica exchange, but also the uniform and well-tempered distributions similarly to metadynamics.
In this sense it provides a unified approach to enhanced sampling.

\section{On-the-fly Probability\\Enhanced Sampling\label{S:opes}}
The recently developed on-the-fly probability enhanced sampling (OPES)\cite{Invernizzi2020} is a collective-variables-based method.
Collective variables (CVs) are function of the microscopic configuration, $\mathbf{s}=\mathbf{s}(\mathbf{x})$, that provide a low dimensional description of the system.
In OPES we aim at modifying the physical probability distribution of $\mathbf{s}$, $P(\mathbf{s})$, in order to reach a given target probability distribution, $p^{tg}(\mathbf{s})$.
To achieve this we must add the following bias potential
\begin{equation}\label{E:opes_bias}
    V(\mathbf{s})=-\frac{1}{\beta}\log \frac{p^{tg}(\mathbf{s})}{P(\mathbf{s})}\, ,
\end{equation}
where $\beta$ is the inverse temperature.
OPES has been introduced as an evolution of metadynamics and in this spirit we first have used the well-tempered target distribution, defined as $p^{WT}(\mathbf{s})\propto [P(\mathbf{s})]^{1/\gamma}$, where $\gamma>1$ is known as bias factor.
This target distribution aims at increasing the transition rate between metastable states of the system, by lowering of a factor $\gamma$ the free energy barriers along the CVs.
In the limit of $\gamma=\infty$ it is equivalent to choosing a uniform target.

Since $P(\mathbf{s})$ is not known a priori, we resort to an iterative scheme.
The core idea in OPES is to estimate the probability distribution at each step $n$, $P_n(\mathbf{s})$, by reweighting on-the-fly a simulation that is biased with $V_n(\mathbf{s})$ which is itself constructed from such $P_n(\mathbf{s})$ estimate according to Eq.~(\ref{E:opes_bias}).
The $P_n(\mathbf{s})$ is obtained via a weighted kernel density estimation, adding a new kernel at a fixed small interval, similarly to metadynamics.

We refer the interested reader to Ref.~\citenum{Invernizzi2020}, where the OPES iterative equations for the case of a well-tempered and a uniform target are presented in detail, and to Refs.~\citenum{Bonati2020, Mandelli2020} for some initial applications.
In the present paper we introduce a class of target distributions that allows sampling any expanded ensemble.
We will also present in detail the OPES iterative scheme for this class of targets.
While the core ingredients of OPES remain the same, the resulting method will look quite different from the one presented in Ref.~\citenum{Invernizzi2020}.
In particular, when targeting expanded ensembles we will not need to use the kernel density estimation that plays instead a crucial role in the well-tempered case.

In applying OPES to sample expanded ensembles we find a method that is similar in spirit to that of Ref.~\citenum{Lyubartsev1992} and to other more recent methods, such as integrated tempering sampling\cite{Gao2008}, infinite switch simulated tempering\cite{Martinsson2019}, and variationally-derived intermediates\cite{Reinhardt2020}.

\section{Targeting Expanded Ensembles\label{S:targeting}}
Let us call $u(\mathbf{x})$ the adimensional reduced potential that contains all the terms depending on the thermodynamic constraints, such as temperature, pressure, or others.
With $\mathbf{x}$ we concisely indicate the atomic coordinates and any other configurational variable that the reduced potential might depend on, such as the volume or the box tensor.
As an example, in the case of the canonical ensemble one has $u(\mathbf{x})=\beta U(\mathbf{x})$, where $\beta$ is the inverse temperature and $U(\mathbf{x})$ is the potential energy of the system.
Let us consider a system with a reduced potential $u_\lambda(\mathbf{x})$ that is a function of $\lambda$, where $\lambda$ could be either a single parameter or a set of parameters, and might indicate e.g.~a thermodynamic property such as the temperature.
At equilibrium its probability distribution follows Boltzmann statistics:
\begin{equation}\label{E:probability}
    P_\lambda(\mathbf{x})=\frac{e^{-u_\lambda(\mathbf{x})}}{Z_\lambda}\, ,
\end{equation}
where $Z_\lambda$ is the partition function, $Z_\lambda=\int e^{-u_\lambda(\mathbf{x})}d \mathbf{x}$.

We are interested in sampling configurations in a range $\Delta \lambda$ of $\lambda$-values.
Instead of running multiple independent simulations at different values of $\lambda$, we can sample a generalized ensemble which contains all the relevant microscopic configurations, and then reweight them to retrieve the correct statistics for any $\lambda\in \Delta \lambda$.
Sampling such ensemble over $\Delta \lambda$ instead of the separate single $\lambda$-ensembles is more efficient when different $\lambda$-ensembles overlap in the coordinate space, and it can also help in solving ergodicity problems.

We must choose as target a distribution that covers all the microscopic configurations relevant to the chosen $\Delta \lambda$ range.
Similarly to what is done in replica exchange, we choose a set $\{\lambda\}$ of $N_{\{\lambda\}}$ values $\lambda \in \Delta \lambda$ such that there is a good overlap between contiguous $P_\lambda(\mathbf{x})$.
We then define our target distribution as:
\begin{equation}\label{E:target}
    p_{\{\lambda\}}(\mathbf{x})=\frac{1}{N_{\{\lambda\}}}\sum_\lambda P_\lambda(\mathbf{x})\, .
\end{equation}
We will refer to this class of target probability distributions as expanded ensemble target distributions.
In the present paper we limit ourselves to considering non-weighted expanded targets, that assign the same $1/N_{\{\lambda\}}$ weight to all the sub-ensembles, but it is also possible to add some $\lambda$-dependent weights and give different importance to different $P_\lambda(\mathbf{x})$.

Without loss of generality, one can consider $\lambda=0$ to be inside the desired interval $\Delta \lambda$.
It is then possible to run a simulation at $\lambda=0$ and use the OPES scheme to iteratively optimize a bias that allows one to sample $p_{\{\lambda\}}(\mathbf{x})$.
Before proceeding to explicitly write the target distribution and the bias potential, we express $P_\lambda(\mathbf{x})$ as
\begin{equation}
    P_\lambda(\mathbf{x})=P_0(\mathbf{x})\, e^{-u_\lambda(\mathbf{x})+u_0(\mathbf{x})}\frac{Z_0}{Z_\lambda}
    =P_0(\mathbf{x})\, e^{-\Delta u_\lambda(\mathbf{x})+\Delta F(\lambda)}\, ,
\end{equation}
where $\Delta u_\lambda(\mathbf{x})=u_\lambda(\mathbf{x})-u_0(\mathbf{x})$ is the potential energy difference and 
\begin{equation}\label{E:deltaF_def}
    \Delta F(\lambda)=-\log \frac{Z_\lambda}{Z_0}
    =-\log \langle e^{-\Delta u_\lambda}\rangle_{u_0}\, ,
\end{equation}
is the dimensionless free energy difference from the reference system $u_0$, that can be expressed also as an ensemble average, indicated with the notation $\langle \cdot \rangle_{u_0}$.
Our expanded target thus becomes
\begin{equation}\label{E:expanded_target}
    p_{\{\lambda\}}(\mathbf{x})=P_0(\mathbf{x})\frac{1}{N_{\{\lambda\}}} \sum_\lambda 
    e^{-\Delta u_\lambda(\mathbf{x})+\Delta F(\lambda)}\, .
\end{equation}

In order to define the target bias, we first rewrite Eq.~(\ref{E:opes_bias}) as
\begin{equation}
    v(\mathbf{x})=-\log \frac{p^{tg}(\mathbf{x})}{P_0(\mathbf{x})}\, .
\end{equation}
Finally the adimensional bias needed to sample the expanded target $p_{\{\lambda\}}(\mathbf{x})$ is:
\begin{equation}\label{E:bias_def}
   v(\mathbf{x})=-\log \left(\frac{1}{N_{\{\lambda\}}}\sum_\lambda e^{-\Delta u_\lambda(\mathbf{x})+\Delta F(\lambda)}\right)\, ,
\end{equation}
that bears some resemblance to the one adopted in parallel bias metadynamics\cite{Pfaendtner2015}.

Note that in writing the bias in this way $P_0(\mathbf{x})$ cancels out.
It follows that the bias potential $v(\mathbf{x})$ depends on the coordinates $\mathbf{x}$ only through the $N_{\{\lambda\}}$ quantities $\Delta u_\lambda (\mathbf{x})$.
We shall refer to these $\Delta u_\lambda$ as expansion collective variables.
The expansion CVs completely characterize the expansion, since not only the bias, but also $\Delta F(\lambda)$, Eq.~(\ref{E:deltaF_def}), and the expanded target distribution $p_{\{\lambda\}}(\mathbf{x})$, Eq.~(\ref{E:expanded_target}), are unambiguously defined through these quantities.
We will see how, by properly choosing the expansion CVs $\Delta u_\lambda (\mathbf{x})$, it is possible to sample different kinds of expanded ensembles.
For each of them we will also highlight the connection between these expansion CVs and more traditional CVs that have a straightforward physical interpretation.

Our target bias, Eq.~(\ref{E:bias_def}), depends on the free energy along the $\lambda$ parameter, $\Delta F(\lambda)$, that is in general unknown.
In the OPES spirit we will reach the target bias iteratively, by estimating on the fly $\Delta F(\lambda)$ via a reweighting procedure, and using such estimate to define the applied bias.

\subsection{Iterative OPES Scheme\label{S:iterative}}
The free energy difference $\Delta F(\lambda)$ defined in Eq.~(\ref{E:deltaF_def}) can be written using an ensemble average over the reference unbiased system $u_0$\cite{Zwanzig1954}.
However, estimating $\langle e^{-\Delta u_\lambda}\rangle_{u_0}$ from an unbiased trajectory is practically impossible due to the extremely small overlap between $P_0$ and $e^{-\Delta u_\lambda}$.
For this reason we use reweighting to write it as an average over the biased ensemble
\begin{equation}\label{E:deltaF_rew}
    e^{-\Delta F(\lambda)}=\langle e^{-\Delta u_\lambda }\rangle_{u_0}=\frac{\langle e^{-\Delta u_\lambda +v} \rangle_{u_0+v}}{\langle e^{v}\rangle_{u_0+v}}\, ,
\end{equation}
where the ensemble average $\langle \cdot \rangle_{u_0+v}$ is computed as a time average over a biased trajectory.
In this way, one can obtain a much more accurate estimate of $\Delta F(\lambda)$.

The problem with this procedure is that the target bias $v$, Eq.~(\ref{E:bias_def}), is itself a function of $\Delta F(\lambda)$.
Therefore we set up an iterative scheme that consists in running a biased simulation whose bias is based on the estimate of the free energy difference that we obtain via on-the-fly reweighting.
At step $n$ the simulation runs with potential $u_0(\mathbf{x})+v_n(\mathbf{x})$, where
\begin{equation}\label{E:bias_n}
    v_n(\mathbf{x})=-\log \left(\frac{1}{N_{\{\lambda\}}}\sum_\lambda e^{-\Delta u_\lambda(\mathbf{x})+\Delta F_n(\lambda)}\right)\, .
\end{equation}
The reweighted estimate $\Delta F_n(\lambda)$ is updated at every iteration step $n$.
In between the iteration steps there is a fixed short stride where the simulation proceeds and both $\Delta F_n(\lambda)$ and the bias $v_n(\mathbf{x})$ are kept constant.
The free energy estimate at the $n$th step can be explicitly written as
\begin{equation}\label{E:deltaF_n}
    \Delta F_n(\lambda)=-\log \left (\frac{\sum\limits_{k=1}^n e^{-\Delta u_\lambda^{(k)}+v_{k-1}^{(k)}}}{\sum\limits_{k=1}^n e^{v_{k-1}^{(k)}}} \right ) \, ,
\end{equation}
where we use the notation $\Delta u_\lambda^{(k)} \equiv \Delta u_\lambda(\mathbf{x}_k)$ and $v_n^{(k)}\equiv v_n(\mathbf{x}_k)$, with $\mathbf{x}_k$ the configuration at the $k$th simulation step.

As the bias approaches convergence, the ensemble sampled approaches the target one, and the $\Delta F_n(\lambda)$ estimates become more and more accurate.
Thus not only do we obtain the target bias, but we also have an estimate of the free energy as a function of the $\lambda$ parameter, i.e. $\Delta F(\lambda)$.
Our iterative scheme is similar in spirit to the one used in integrated tempering sampling\cite{Gao2008}, but the two differ both in their implementation and in their applications.

Eqs.~(\ref{E:bias_n}) and (\ref{E:deltaF_n}) are the explicit OPES iterative equations used for sampling the expanded ensemble defined by the target distribution $p_{\{\lambda\}}(\mathbf{x})$, Eq.~(\ref{E:target}), and are at the heart of our method.
In the following sections we will show how these equations can be used to sample different expanded ensembles, simply by specifying different expansion CVs $\Delta u_\lambda(\mathbf{x})$.
Once these are chosen, the only free parameter of the method is the stride between the iterations.
This should be set so that consecutive steps are not too correlated, as it is the case for the on-the-fly Gaussians deposition in metadynamics.

It is possible to parallelize the procedure using multiple replicas of the system, as done in  multiple walkers metadynamics\cite{Raiteri2006}, where each replica shares the same bias and all contribute to the ensemble average in Eq.~(\ref{E:deltaF_n}).
At variance with replica exchange, the number of parallel simulations does not have to be equal to the number $N_{\{\lambda\}}$ of $\lambda$-points that define the target.

Finally we notice that one could consider expressing the free energy $\Delta F(\lambda)$ via a cumulant expansion\cite{Park2008,Radak2018}.
This generally provides a very good estimate close to the reference $\lambda=0$, but can be very inaccurate when the range is broad, requiring a great number of terms in the expansion. 
Furthermore we found it can introduce artificial barriers that might stop the system from visiting all the relevant configurations, thus making the OPES self-consistent procedure much less efficient.

\subsection{Reweighting\label{S:reweighting}}
Until now we have seen how to sample expanded ensembles by applying a bias potential.
We now need a reweighting procedure in order to retrieve statistics at any desired value of $\lambda$.
To this effect one can use standard umbrella sampling reweighting\cite{Torrie1977}.
Given any observable $O=O(\mathbf{x})$ that is a function of the atomic coordinates, we can calculate its average in the ensemble $\lambda$ via the following reweighting equation:
\begin{equation}\label{E:reweight}
    \langle O \rangle_{u_\lambda}=\frac{\langle O e^{-\Delta u_\lambda+v}\rangle_{u_0+v}}{\langle e^{-\Delta u_\lambda+v}\rangle_{u_0+v}}
    \approx \frac{\sum_k^n O_k w_k(\lambda)}{\sum_k^n w_k(\lambda)}\, ,
\end{equation}
where $O_k\equiv O(\mathbf{x}_k)$ and the weight $w_k(\lambda)$ is defined as $w_k(\lambda)\equiv e^{-\Delta u_\lambda^{(k)}+v_{k-1}^{(k)}}$.

This equation assumes that the applied bias is static or quasi-static, meaning that it is updated in an adiabatic fashion.
It is thus good practice to discard an initial transient of the simulation, where the bias changes quickly, and not use it for reweighting.
Determining the exact length to be discarded might not be intuitive, however OPES generally assigns a very low weight to this initial transient, and thus the result will not be significantly affected by this choice\cite{Invernizzi2020}.

A useful diagnostic tool when performing reweighting is the so-called effective sample size, defined as the number of sampled points $n$ times the ratio between the variance of an observable in the unbiased ensemble and its variance in the reweighted ensemble\cite{Kong1994}.
For practical purposes we will use not this definition, but rather the popular estimator\cite{Kish1965} defined as:
\begin{equation}\label{E:neff}
    n_{\text{eff}}(\lambda)=\frac{\left[\sum_k w_k(\lambda)\right]^2}{\sum_k w_k^2(\lambda)}\, ,
\end{equation}
where $w_k(\lambda)$ are the importance sampling weights.
Intuitively, the effective sample size for a given $\lambda$ will be smaller than the total number of samples, $n_{\text{eff}}(\lambda)<n$.
One should expect the efficiency to be roughly $n_{\text{eff}}(\lambda)/n\propto 1/N_{\{\lambda\}}$, given a minimal choice of $\lambda$-points that properly covers the target range.
Plotting $n_{\text{eff}}/n$ as a function of $\lambda$ can be a good diagnostic tool to monitor the consistency of the iterative procedure.

Finally we notice that the estimate of uncertainties requires some extra care in case of weighted samples\cite{Bussi2019}.
In the Appendix~\ref{A:block} we describe in detail the weighted block averaging procedure we adopt, and show how the effective sample size plays a role.

\section{Linearly Expanded Ensembles\label{S:linearly}}
An important type of expanded ensemble is the one obtained by linearly perturbing the reduced potential of the system, $u_\lambda(\mathbf{x})=u_0(\mathbf{x})+\lambda \Delta u(\mathbf{x})$.
It is defined by the following expansion CVs
\begin{equation}
    \Delta u_\lambda(\mathbf{x})=\lambda \Delta u(\mathbf{x})\, .
\end{equation}

Various different ensembles can be obtained in this way, such as the multicanonical ensemble and the multibaric ensemble, but also alchemical transformations, and others.
Recently an interesting ``multiforce'' ensemble that falls in this category has been proposed\cite{Hartmann2020}.
We will present in detail some of these ensembles in the following sections.

It can be useful to group together these linearly expanded ensembles because they share some interesting properties.
In particular for these ensembles we can propose a simple automatic way to chose the $\lambda$-points that define the target $p_{\{\lambda\}}(\mathbf{x})$.
The idea is to have the $\lambda$-points uniformly distributed in the $\Delta \lambda$ interval with a spacing $\Delta \lambda/N_{\{\lambda\}}$ estimated from the effective sample size as a function of $\lambda$, $n_{\text{eff}}(\lambda)$.
In practice what we do is to run a short unbiased simulation of $n$ steps at $\lambda=0$ and use a root finding algorithm to determine the points $\lambda_+>0$ and $\lambda_-<0$ such that  $n_{\text{eff}}(\lambda_\pm)/n \approx 0.5$.
Then one can use a total of $N_{\{\lambda\}}=\Delta \lambda/(\lambda_++\lambda_-)$ equally spaced $\lambda$-points to define the target $p_{\{\lambda\}}(\mathbf{x})$.

This heuristic way of choosing the $\lambda$-points is not optimal and more elaborate options have been explored in the replica exchange literature\cite{Prakash2011}.
However, in our case this choice is less critical, since within our scheme one can increase $N_{\{\lambda\}}$ without the need to simulate additional replicas of the system.
Thus this procedure provides an easy and automatic guess for linearly expanded ensembles that can be practically useful in many scenarios.

\subsection{Multicanonical Ensemble\label{S:multicanonical}}
We start by considering as example of linearly expanded ensemble the case of the multicanonical ensemble, which is probably the one with the longest history.
The goal is to sample all the configurations relevant for canonical simulations in a given range of temperatures.
In a canonical simulation the reduced potential is $u(\mathbf{x})=\beta U(\mathbf{x})$, where $U(\mathbf{x})$ is the potential energy, $\beta=1/(k_B T)$ is the inverse thermodynamic temperature and $k_B$ is Boltzmann constant.
It is possible to define a multicanonical linearly expanded ensemble, by putting $\Delta u(\mathbf{x})=u_0(\mathbf{x})=\beta_0U(\mathbf{x})$ and $\lambda=(\beta-\beta_0)/\beta_0$, where $\beta_0$ is the inverse temperature set by the thermostat of the simulation, and $\beta$ spans the target range, $\beta_{\min}<\beta<\beta_{\max}$.

The expansion CVs that define such target are 
\begin{equation}
    \Delta u_\lambda(\mathbf{x})=\lambda \beta_0U(\mathbf{x})=(\beta-\beta_0)U(\mathbf{x})\, ,
\end{equation}
and by using them in the OPES iterative equations, Eqs.~(\ref{E:bias_n}) and (\ref{E:deltaF_n}), we obtain our multicanonical simulation.
Given the physical significance of the inverse temperature $\beta$, it is more natural to directly consider $\beta$ as parameter instead of the dimensionless $\lambda$.
We thus write $\Delta u_\beta$ and $\Delta F=\Delta F(\beta)$, where we have set $\Delta F(\beta_0)=0$.
Similarly, it is natural to consider the potential energy $U(\mathbf{x})$ as collective variable, and thus write the bias as
\begin{equation}
    v(U)=-\log \left( \frac{1}{N_{\{\beta\}}}\sum_\beta e^{-(\beta-\beta_0)U + \Delta F(\beta)}\right)\, .
\end{equation}

It is important to notice that we did not require the bias to be a function of a single CV, but rather we find it to be the case when we set as target the temperature-expanded ensemble.
This is in fact a general property of linearly expanded ensembles.
When expanding according to a given $\lambda$, the resulting bias will be a function only of the thermodynamic conjugate variable $\Delta u$.
To define the bias $v=v(\Delta u)$ we then need to estimate the free energy along $\lambda$, $\Delta F(\lambda)$.

Other multicanonical methods aim instead at sampling a flat energy distribution\cite{Berg1991,Lee1993,Piaggi2019}.
In order to do so, they need to estimate the free energy as a function of $U$ (or equivalently the density of states along $U$), while in our method, as in other tempering approaches\cite{Martinsson2019}, we instead need to estimate the free energy as a function of temperature, $\Delta F(\beta)$.

\subsubsection*{Example: Alanine Dipeptide}
As an example of multicanonical sampling we consider alanine dipeptide in a vacuum, at temperature $T_0=300$~K.
This is a typical toy model for testing enhanced sampling methods, since at room temperature it presents two metastable states with an extremely low transition probability.
A possible way of enhancing the sampling is to bias the $\phi$ and $\psi$ dihedral angles, using as target a flat uniform distribution or the well-tempered distribution\cite{Invernizzi2020}.
Here instead we bias the potential energy $U$, and use as a target the multicanonical ensemble over a temperature range from 300~K up to 1000~K.

Simulations are performed with the molecular dynamics software GROMACS\cite{gromacs}, patched with the enhanced sampling library PLUMED\cite{plumed} (see \href{https://arxiv.org/src/2007.03055v3/anc/SupplementalMaterial.pdf}{SM} for computational details).
The only input needed for OPES, beside the temperature range we are interested in sampling, is the stride with which updating the bias, that is taken to be 500 simulation steps (1~ps).

\begin{figure*}
  \begin{tabular}{cc}
        \includegraphics[width=0.45\textwidth]{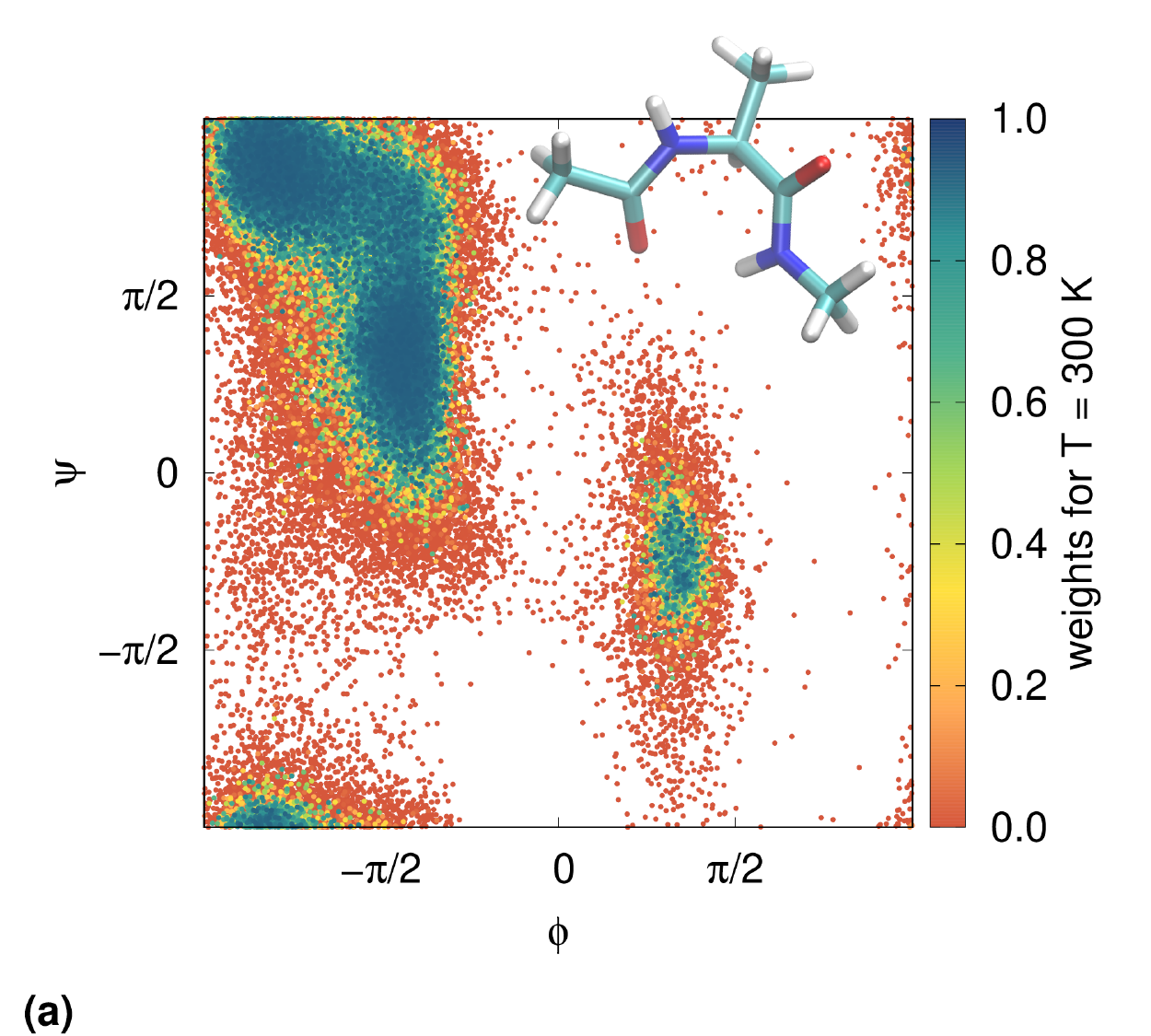}
        &  \includegraphics[width=0.45\textwidth]{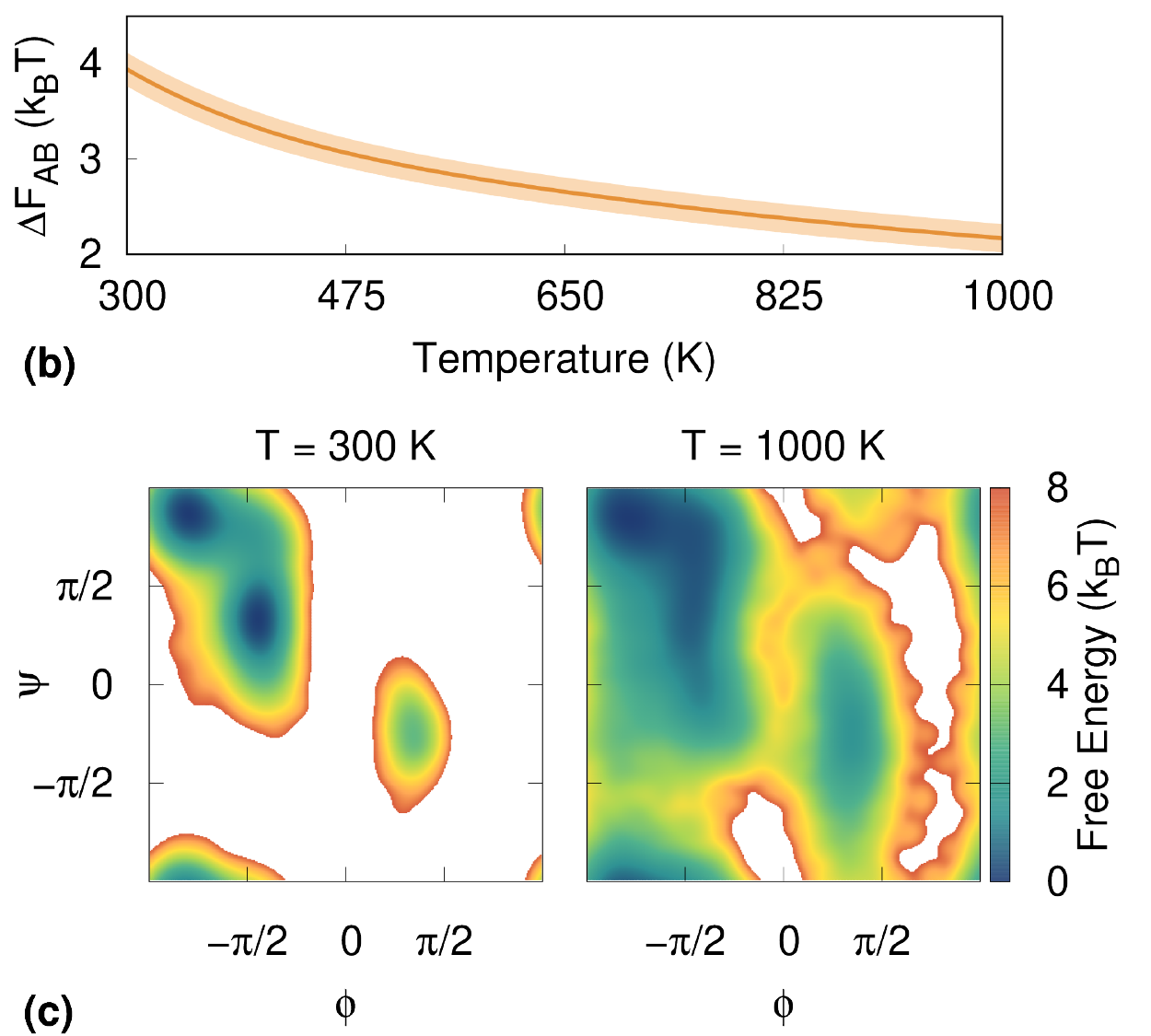}
  \end{tabular}
  \caption{Alanine dipeptide in the multicanonical ensemble ($T_{\min}= 300$~K, $T_{\max}=1000$~K). 
  (a) Explored configurations as a function of the dihedral angles.
  Sampled points are colored according to their reweighting weight at $T=T_0=300$ K, $w_k(\beta)=e^{-(\beta-\beta_0)U_k+v_{k-1}^{(k)}}$.
  Notice how all the points in the transition state, close to $\phi=0$, have extremely low probability of being sampled in an isothermal simulation at $300$~K.
  (b) Free energy difference between the two basins $\Delta F_{AB}$ as a function of temperature.
  (c) Reweighted free energy surface at two different temperatures.
  \label{F:ala2}}
\end{figure*}
Fig.~\ref{F:ala2}a shows on the $\phi,\psi$ plane the configurations sampled during the 100~ns multicanonical run.
It is interesting to notice that the potential energy $U$ would be considered a bad CV in enhanced sampling methods such as metadynamics, since it cannot distinguish between the two basins, that have roughly the same energy.
However, when using the multicanonical ensemble as target, by biasing $U$ we can enhance the probability of visiting the transition state (roughly the region where $\phi=0$), and thus observe multiple transitions between the basins and converge the free energy difference between them, $\Delta F_{AB}$ (Fig.~\ref{F:ala2}b).
We can use the angle $\phi$ to define this free energy difference between the two basins:
\begin{equation}
  \Delta F_{AB}=-\log \left(\frac{\langle \chi_{\phi\in[0,\pi]} \rangle}{\langle \chi_{\phi\in[-\pi,0]} \rangle}\right)\, ,
\end{equation}
where $\chi$ is a characteristic function, equal to 1 if the variable is in the proper range and 0 otherwise.

In the \href{https://arxiv.org/src/2007.03055v3/anc/SupplementalMaterial.pdf}{supplemental material} we show a comparison between this multicanonical run and a well-tempered run biasing the two dihedral angles.
As expected the latter is much more efficient (roughly 10 times) in converging the free energy difference at a single temperature, due to the fact that it focuses on the relevant degrees of freedom.

\subsection{Multithermal-Multibaric Ensemble\label{S:multiTP}}
Within our scheme, combining different linearly expanded ensembles is straightforward.
One simply has a two dimensional parameter $\lambda=\{\lambda_1,\lambda_2\}$, and considers $u_\lambda(\mathbf{x})=u_0+\lambda_1\Delta u_1(\mathbf{x})+\lambda_2\Delta u_2(\mathbf{x})$.
This can be useful for example to sample multiple temperatures and multiple pressures in a single multithermal-multibaric simulation.

In this case we consider NPT simulations with a reference reduced potential $u_0(\mathbf{x})=\beta_0 U(\mathbf{x})+\beta_0 p_0 V(\mathbf{x})$, where $p$ is the pressure and $V(\mathbf{x})$ the volume.
Similarly to what done before, it is more natural to use as $\lambda$ parameters directly the temperature $\beta$ and the pressure $p$, and write the expansion CVs $\Delta u_\lambda(\mathbf{x})$ as 
\begin{equation}\label{E:multitp_ecv}
    \Delta u_{\beta,p}(\mathbf{x})=(\beta-\beta_0) U(\mathbf{x})+(\beta p -\beta_0 p_0)V(\mathbf{x})\, .
\end{equation}
The target distribution is defined by a set of $N_{\{\beta\}}$ temperatures $\beta \in [\beta_{\min},\beta_{\max}]$ and $N_{\{p\}}$ pressures $p \in [p_{\min},p_{\max}]$, for a total of $N_{\{\beta,p\}}=N_{\{\beta\}} \times N_{\{p\}}$ different $\Delta F(\beta,p)$ to be estimated.
We will also express the bias, Eq.~(\ref{E:bias_def}), as a function of the potential energy and the volume $v=v(U,V)$, which come as a natural CVs choice.
As already discussed, the intermediate temperatures $\beta$ and pressures $p$ that define the target can be chosen automatically from a short unbiased simulation.
We can do this independently for the two parameters, despite the fact that the pressure term $p$ is multiplied by $\beta$ in Eq.~\ref{E:multitp_ecv}.

Finally we notice how the choice of $\beta_0$ and $p_0$ is completely free.
As long as they lay inside the range of temperatures and pressures that we aim at sampling, no matter what thermodynamic conditions we start from at convergence we will sample the same configurational space.
However, when the target range is very broad, choosing $\beta_0$ and $p_0$ roughly at the center can help to speed up convergence.

\subsubsection*{Example: Chignolin}
As an example of a multithermal-multibaric simulation we consider the miniprotein chignolin (CLN025) with CHARMM22* force field\cite{Piana2011} and TIP3P water (about 2800 molecules), over a temperature range from $T_{\min}=270$~K to $T_{\max}=800$~K and a pressure range from $p_{\min}=1$~bar to $p_{\max}=4000$ bar.
The velocity-rescaled thermostat\cite{Bussi2007} is set at $T_0=500$~K and the Parrinello-Rahman barostat\cite{Parrinello1981} at $p_0=2000$~bar.
The $\Delta F_n(\beta,p)$ estimates and the bias are updated every 500 simulation steps (1~ps).
The $N_{\{\beta\}}$ temperature steps and $N_{\{p\}}$ pressure steps are chosen automatically based on a short 100~ps unbiased run.
This results in 92 steps in temperature and 26 in  pressure, for a total of $N_{\{\beta,p\}}=2392$ points.
In order to avoid the region of low pressure and high temperature where water could evaporate, we discard any $(\beta,p)$-point laying below the line from (500~K, 1~bar) and (800~K, 1000~bar).
In this way 91 $(\beta,p)$-points are discarded.

The simulation is performed in parallel using 40 multiple walkers, and runs for a total of 300~ns of which roughly 10~ns are needed to converge the bias.
Also in this case we use GROMACS patched with PLUMED.

\begin{figure*}
    \begin{tabular}{cc}
        \includegraphics[width=0.35\textwidth]{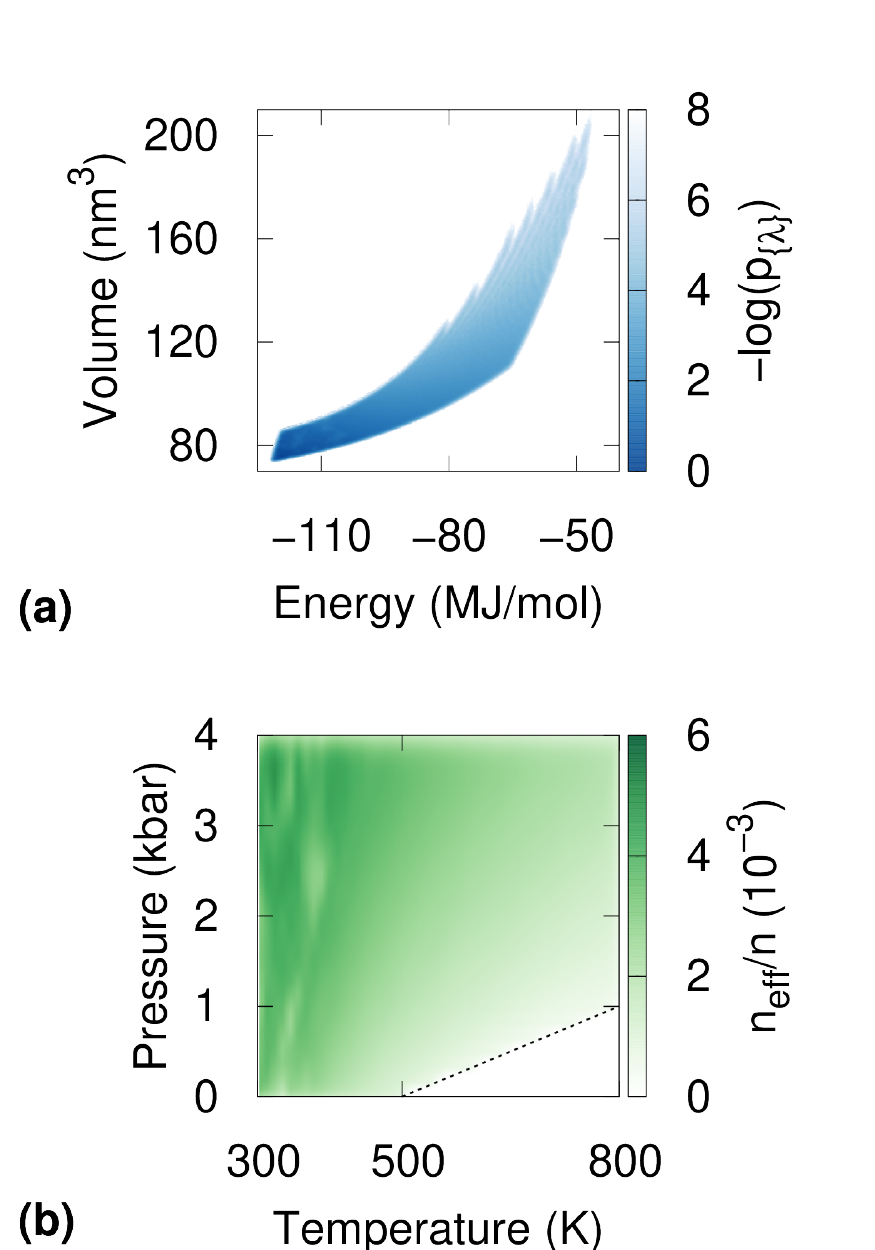}
        & \includegraphics[width=0.6\textwidth]{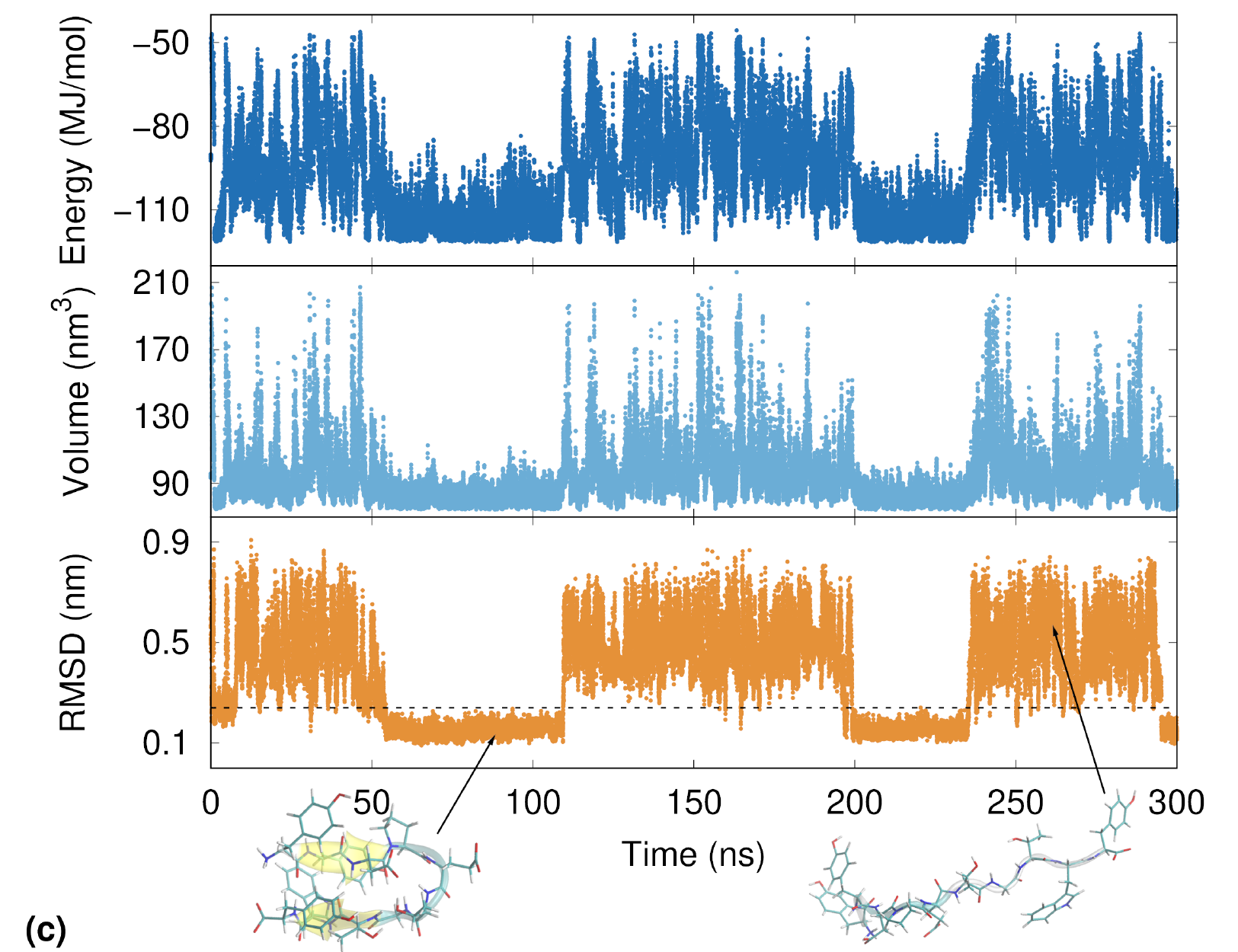}
    \end{tabular}
    \caption{Multithermal-multibaric simulation of chignolin.
    (a) Sampled target distribution in the CV space of potential energy and volume.
    (b) Relative effective sample size at different temperatures and pressures.
    The bottom corner of high temperatures and low pressures is excluded from the target to avoid vaporization of the system.
    (c) Time evolution of the two biased CVs and of the C$\alpha$-RSMD for one of the 40 walkers.
    An RMSD threshold between folded and unfolded is highlighted with a dashed line.
    \label{F:chignolin}}
\end{figure*}

In Fig.~\ref{F:chignolin}a we show the distribution sampled in the energy-volume space, while in Fig.~\ref{F:chignolin}b the corresponding effective sample size $n_{\text{eff}}$ is plotted, as a function of temperature and pressure and rescaled over the total number of samples $n$.
The $n_{\text{eff}}/n$ is not perfectly uniform, but it has the same order of magnitude over the whole target region. 
On the right, Fig.~\ref{F:chignolin}c, we show for one of the 40 replicas the energy and volume trajectory, together with the trajectory of the C$\alpha$-RMSD to the experimental NMR structure\cite{Honda2008}.

\begin{figure}
    \includegraphics[width=0.45\textwidth]{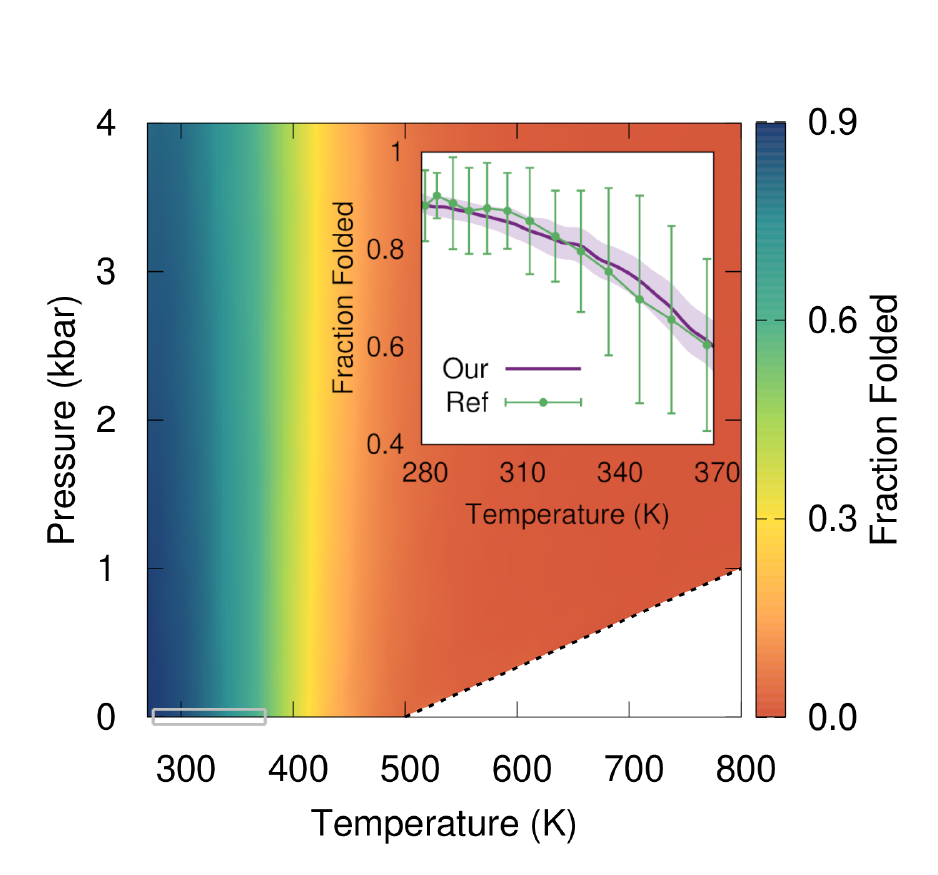}
    \caption{The fraction folded of chignolin estimated from the multithermal-multibaric simulation.
    The inset shows the same quantity over a smaller range of temperatures at 1~bar (highlighted with a gray box), in order to compare it with the reference results from Lindorff-Larsen et al.\cite{Lindorff-Larsen2012}.
    \label{F:chigno-folding}}
\end{figure}
In Fig.~\ref{F:chigno-folding} we show the chignolin folded fraction at the different temperatures and pressures we targeted.
The folded fraction is defined as in Ref.~\citenum{Lindorff-Larsen2012}, using a dual-cutoff on the C$\alpha$-RMSD based on the CLN025 experimental NMR structure.
A configuration is considered folded when the RMSD goes below 0.1 nm, and unfolded when it goes above 0.4 nm.
In the inset we compare our results with those of Ref.~\citenum{Lindorff-Larsen2012}, that considered a smaller temperature range at standard pressure.
The confidence interval of our estimate is calculated with the block analysis described in Appendix~\ref{A:block}.

The stability diagram of chignolin (Fig.~\ref{F:chigno-folding}) does not present striking features, in qualitative agreement with Ref.~\citenum{Okumura2012}.
However, it has been recently shown\cite{Uralcan2019} that other miniprotein can have a non-trivial phase diagram, with unfolding both at low temperature and at high pressure.

\subsection{Thermodynamic Integration\label{S:termoint}}
Another interesting application of our method is its use for performing thermodynamic integration\cite{FrenkelBook}. 
Let's consider a system with reduced potential energy $u_0(\mathbf{x})$ and free energy $F_0=-\log Z_0$ and another similar system with potential $u_1(\mathbf{x})$ and free energy $F_1$. 
We are interested in calculating the free energy difference $\Delta F_{0\rightarrow1}=F_1-F_0$, for instance because we know the free energy of one of the two systems and in this way we can retrieve the other one.
The key idea of thermodynamic integration is to define a ladder of intermediate systems with reduced potentials $u_\lambda(\mathbf{x})$ and $0<\lambda <1$, to connect the two systems.
The free energy difference $\Delta F_{0\rightarrow1}$ to go from the $u_0$ system to the $u_1$ can be calculated via the following integral:
\begin{equation}\label{eq:thermo_integration}
    \Delta F_{0\rightarrow1}=\int\limits_0^1 \left \langle \frac{\partial u_{\lambda}(\mathbf{x})}{\partial \lambda} \right \rangle_{u_\lambda}d\lambda\, .
\end{equation}

Typically individual simulations are run using $u_{\lambda}(\mathbf{x})$ for different values of $\lambda$ and the ensemble average $\left \langle \frac{\partial u_{\lambda}(\mathbf{x})}{\partial \lambda} \right \rangle_{u_\lambda}$ is estimated for each of them. 
Then the integration in Eq.~(\ref{eq:thermo_integration}) can be carried out numerically, e.g.~using the trapezoid rule or a Gaussian quadrature.

The most common way to define the intermediate states $u_\lambda(\mathbf{x})$ is via a linear interpolation
\begin{equation}\label{E:thermo_interpolation}
    u_\lambda(\mathbf{x})=u_0(\mathbf{x})+\lambda \Delta u(\mathbf{x})\, ,
\end{equation}
where $\Delta u(\mathbf{x})\equiv u_1(\mathbf{x})-u_0(\mathbf{x})$.
In this case we have $\partial u_\lambda/\partial \lambda=\Delta u$.

In the spirit of the present paper, we aim at performing a single simulation that samples all values of $\lambda$ simultaneously.
It is then possible to reweight for any $\lambda$ and calculate the integral in Eq.~(\ref{eq:thermo_integration}).
Thus we simulate the system at $u_0(\mathbf{x})$ and build a target $p_{\{\lambda\}}(\mathbf{x})$ as in Eq.~(\ref{E:target}) using $N_{\{\lambda\}}$ $\lambda$-points in the interval $0<\lambda<1$.
The OPES iterative equations, Eqs.~(\ref{E:bias_n}) and (\ref{E:deltaF_n}), can be written using the expansion CVs $\Delta u_\lambda=\lambda \Delta u$ as defined in Eq.~(\ref{E:thermo_interpolation}).

Finally we notice that thermodynamic integration can be performed using interpolation schemes different from the linear one, and our method is general and can be applied also in those scenarios, simply by properly defining the expansion CVs $\Delta u_\lambda(\mathbf{x})$.

\subsubsection*{Example: TIP4P Water to Lennard-Jones}
We now use the thermodynamic integration formalism described above to calculate the free energy of TIP4P water, relative to a reference Lennard-Jones system.
The TIP4P potential energy ($U_{TIP4P}$) is made of an electrostatic energy term and a van der Waals type interaction between the oxygens described by a Lenard-Jones potential ($U_{LJ}$).
The free energy of a Lennard-Jones fluid with the same $U_{LJ}$ potential has been fit to an equation of state and thus is a good reference\cite{Johnson1993}.
For the simulations we use the LAMMPS\cite{lammps} molecular dynamics software, patched with PLUMED.
We perform an NVT canonical simulation at 443~K using the TIP4P water potential, thus $u_0(\mathbf{x})=\beta U_{TIP4P}(\mathbf{x})$, with $N=384$ molecules.
The reference system is characterized by $u_1(\mathbf{x})=\beta U_{LJ}(\mathbf{x})$.

\begin{figure*}
    \begin{tabular}{cc}
        \includegraphics[width=0.45\textwidth]{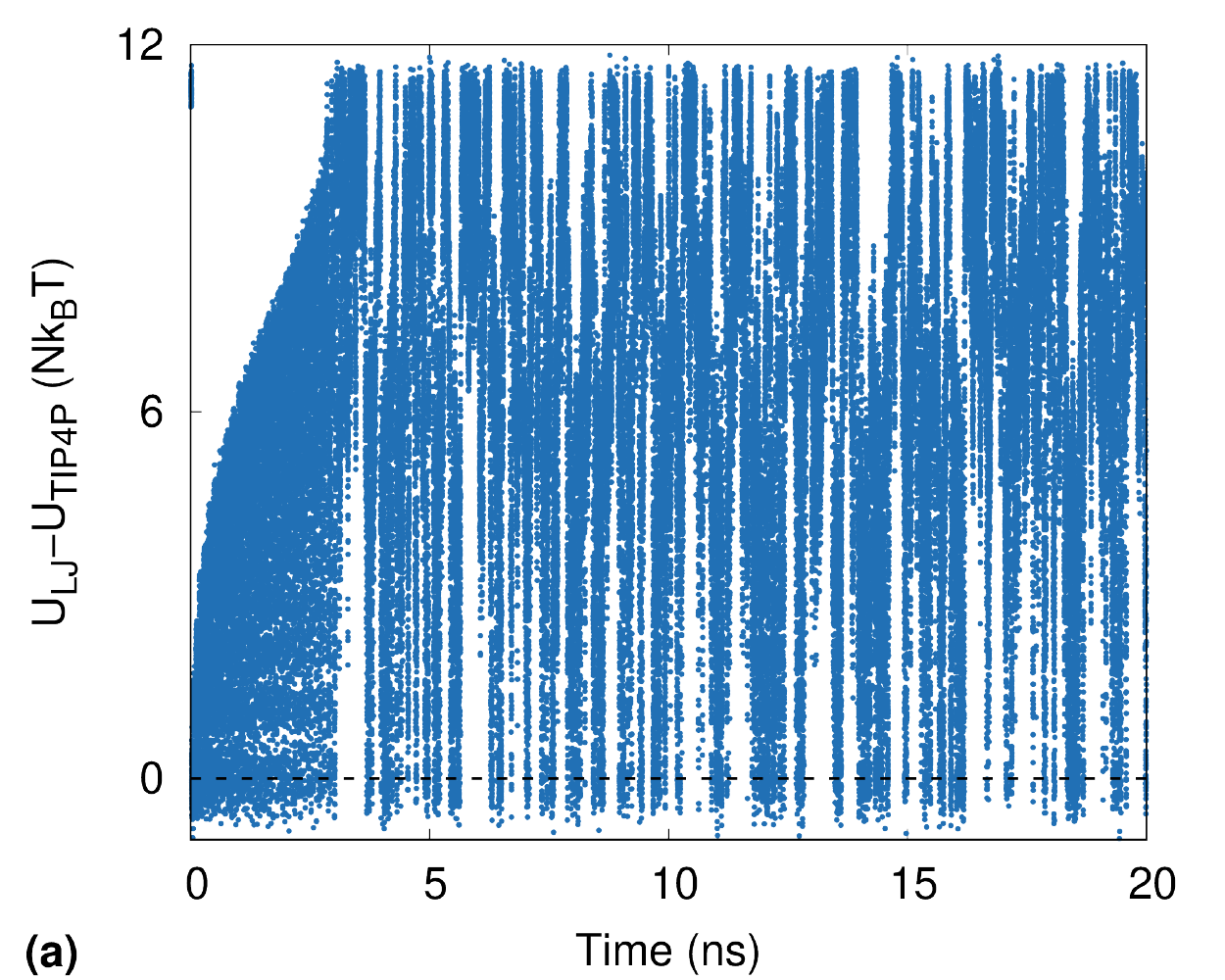}
        & \includegraphics[width=0.45\textwidth]{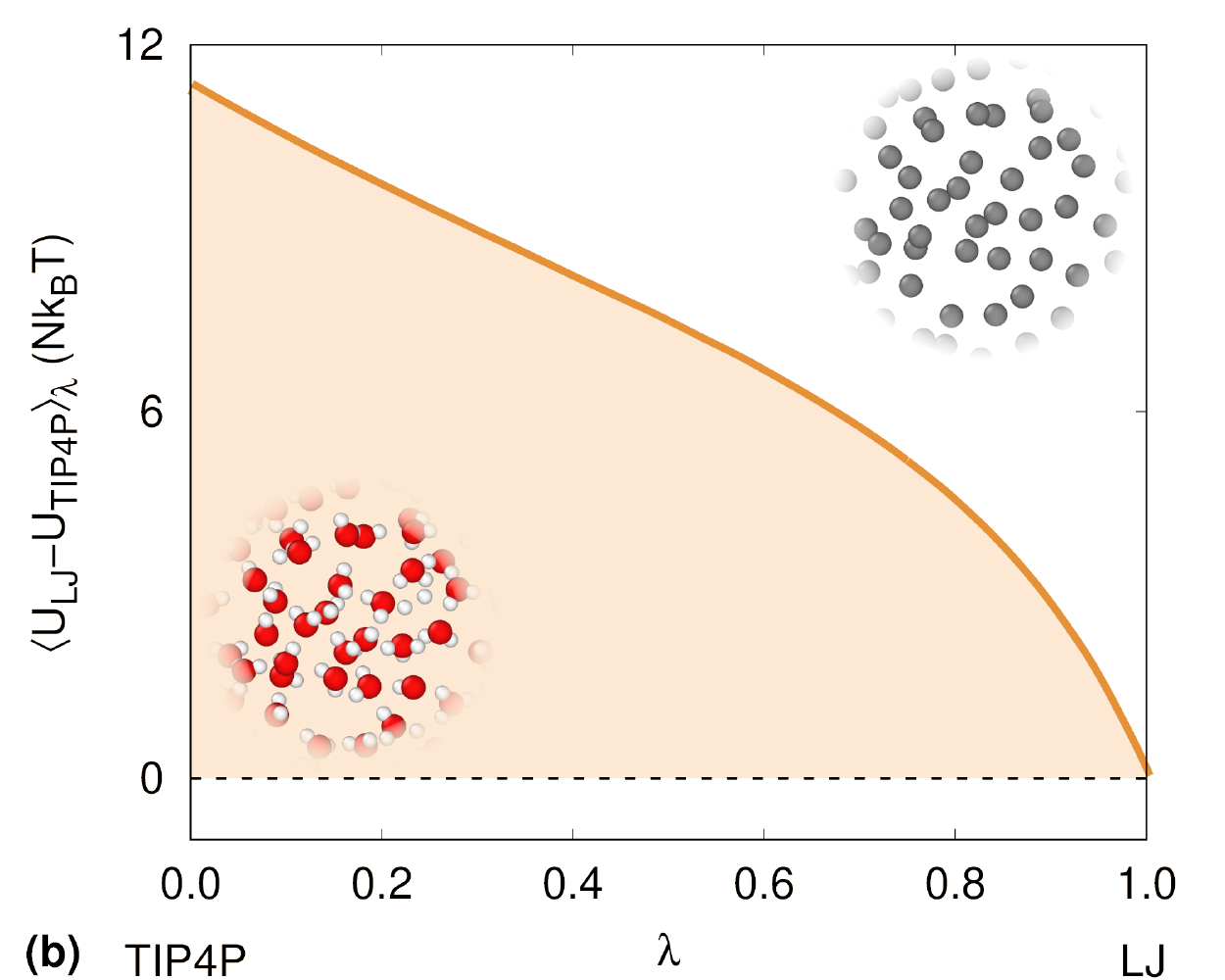}
    \end{tabular}
    \caption{Calculation of the free energy of liquid TIP4P water using a single-simulation thermodynamic integration.
    (a) Evolution of the collective variable $\Delta U$ as a function of simulation time.
    (b) Integrand for the thermodynamic integration obtained through reweighting.
    \label{F:multilambda}}
\end{figure*}

Being $\beta$ a constant, we consider as collective variable $\Delta U(\mathbf{x})\equiv U_{LJ}(\mathbf{x})-U_{TIP4P}(\mathbf{x})$, and write the bias according to Eq.~(\ref{E:bias_def}):
\begin{equation}
    v(\Delta U)=-\log \left( \frac{1}{N_{\{\lambda\}}}\sum_\lambda e^{-\lambda\beta\Delta U + \Delta F(\lambda)}\right)\, .
\end{equation}
From a short 20~ps unbiased run we obtain with the usual automatic procedure (Sec.~\ref{S:linearly}) 30 equispaced points in the interval $0<\lambda<1$, that define our target distribution.
The evolution of $\Delta U$ as a function of simulation time is shown in Fig.~\ref{F:multilambda}a.
There is an initial transient of about 3~ns until the bias potential is optimized and then the system diffuses freely.
This has to be compared with a simulation for a given value of $\lambda$ in which the fluctuations of $\Delta U$ would be very small.
From this simulation the integrand $\left \langle \frac{\partial u_{\lambda}(\mathbf{x})}{\partial \lambda} \right \rangle_{u_\lambda} = \beta \langle \Delta U \rangle_{\lambda}$ can be calculated via reweighting, Eq.~(\ref{E:reweight}),
\begin{equation}
    \langle \Delta U \rangle_{\lambda}=
    \frac{\sum_k^n \Delta U_k \, w_k(\lambda)}{\sum_k^n w_k(\lambda)}\, ,
\end{equation}
where $\Delta U_k=U_{LJ}(\mathbf{x}_k)-U_{TIP4P}(\mathbf{x}_k)$ and $w_k(\lambda)=e^{-\lambda \beta \Delta U_k+v_{k-1}^{(k)}}$.
The values of $\langle \Delta U \rangle_{\lambda}$ thus calculated are shown in Fig.~\ref{F:multilambda}b.
Using these results and Eq.~(\ref{eq:thermo_integration}) we find a free energy difference $\Delta F_{TIP4P\rightarrow LJ}=F_{LJ}-F_{TIP4P}=7.00(1)$ ($Nk_BT$ units) in agreement with the result reported in Ref.~\citenum{Vega2008}.

%

\section{Beyond linearity:\\Multiumbrella Ensemble\label{S:multiumbrellas}}
We consider now another important kind of expanded ensemble, namely the one obtained by combining all the different windows of a typical umbrella sampling simulation\cite{Sugita2000}.
We will refer to such an ensemble as multiumbrella ensemble.

Multiple windows umbrella sampling\cite{Kastner2011} allows for the free energy surface (FES) reconstruction along some collective variable $s=s(\mathbf{x})$, that can be the reaction coordinate or some slow mode of the system.
Typically one simulates multiple copies of the system, each one with a parabolic bias potential centered at a different $s_\lambda$-point, in such a way that the resulting probability distributions have an overlap and cover the whole CV range.
Post-processing via WHAM\cite{Kumar1992} or other methods is then needed to combine the data in a single FES estimate.
Here instead we aim at sampling all the umbrella windows in the same simulation via a single global potential, and obtain the FES with the simple reweighting scheme described in Sec~\ref{S:reweighting}, without further  post-processing.

Given a system with reduced potential $u_0(\mathbf{x})$ and equilibrium Boltzmann distribution $P_0(\mathbf{x})$, we can write the reduced potential of each umbrella window as $u_\lambda(\mathbf{x})=u_0(\mathbf{x})+\Delta u_\lambda(\mathbf{x})$, with expansion CVs
\begin{equation}\label{E:umbrella_ecv}
    \Delta u_\lambda(\mathbf{x})=\frac{(s(\mathbf{x})-s_\lambda)^2}{2\sigma^2}\, .
\end{equation}
The associated probability distribution is $P_\lambda(\mathbf{x}) \propto P_0(\mathbf{x})G_\sigma(s(\mathbf{x}),s_\lambda)$, where $G_\sigma(s,s_\lambda)$ is a Gaussian of width $\sigma$ centered in $s_\lambda$.
The resulting expanded target $p_{\{\lambda\}}(\mathbf{x})=\frac{1}{N_{\{\lambda\}}}\sum_\lambda P_\lambda(\mathbf{x})$ is clearly not linear in $\lambda$, and in fact requires an extra parameter $\sigma$ to be defined.
The width $\sigma$ can in principle vary with $\lambda$, but we consider here only the case of uniform umbrellas.

Since the expansion CVs, Eq.~(\ref{E:umbrella_ecv}), depend on $\mathbf{x}$ only through $s=s(\mathbf{x})$, it is natural to write the bias as a function of the $s$ CV
\begin{equation}
    v(s)=-\log \left( \frac{1}{N_{\{\lambda\}}}\sum_\lambda e^{-\frac{(s-s_\lambda)^2}{2\sigma^2}+ \Delta F(s_\lambda)} \right)\, .
\end{equation}
Contrary to the linear case, in this multiumbrella case both the bias $v(s)$ and the free energy differences ${\Delta F(s)=-\log \langle G_\sigma(s(\mathbf{x}),s)\rangle_{u_0}}$ are expressed as functions of the same CV.

The $N_{\{\lambda\}}$ $s_\lambda$-points can be chosen to be uniformly distributed in the desired $\Delta s=s_{\max}-s_{\min}$ interval, in such a way to be at most at a distance of $\sigma$, ensuring overlap between contiguous $P_\lambda$.
For a small enough $\sigma$, the estimate $\Delta F_n(s)$ converges precisely to the free energy surface (FES), while if $\sigma$ is too broad there will be small artifacts, similarly to what happens when a too broad bandwidth is used in kernel density estimation.

It is instructive to consider the marginal of the target probability with respect to the CV, $p_{\{\lambda\}}(s)$.
In the limit of infinitely small $\sigma$ and thus infinitely large $N_{\{\lambda\}}$, the multiumbrella target $p_{\{\lambda\}}(s)$ is a uniform flat distribution over the $\Delta s$ interval.
In the opposite limit, of a very broad $\sigma$, the target distribution will look like the original, hard-to-sample $P_0$.
As a rule of thumb $\sigma$ should be as small as the smallest features of the FES we are interested in.
We notice that this is the same criterion used to chose the $\sigma$ parameter in metadynamics\cite{Valsson2016}, and it can typically be guessed from a short unbiased run.
For this reason we prefer to use as parameter $\sigma$ instead of the more commonly used strength of the harmonic umbrella potential $K=1/\sigma^2$\cite{Kastner2011}.

In some cases it proved useful to introduce two small modifications to make the multiumbrella iterative optimization scheme more robust.
We leave the explanation of them to Appendix~\ref{A:robustness}, since they have not been necessary for the examples presented in the paper.

For simplifying the exposition we presented the procedure in case of a 1D CV, but it is straightforward to extend it to higher dimension, by using multidimensional Gaussians and placing the $s_\lambda$-points on an appropriate multidimensional grid.
When dealing with higher dimensions it might be interesting to use some more elaborate shapes for the umbrellas, e.g.~a Gaussian mixture in a similar but complementary way to Ref.~\citenum{Debnath2020}, or to follow a specific path, as in Ref.~\citenum{Branduardi2007}.

\subsubsection*{Example: Double-well Model}
\begin{figure*}
    \begin{tabular}{cc}
        \includegraphics[width=0.45\textwidth]{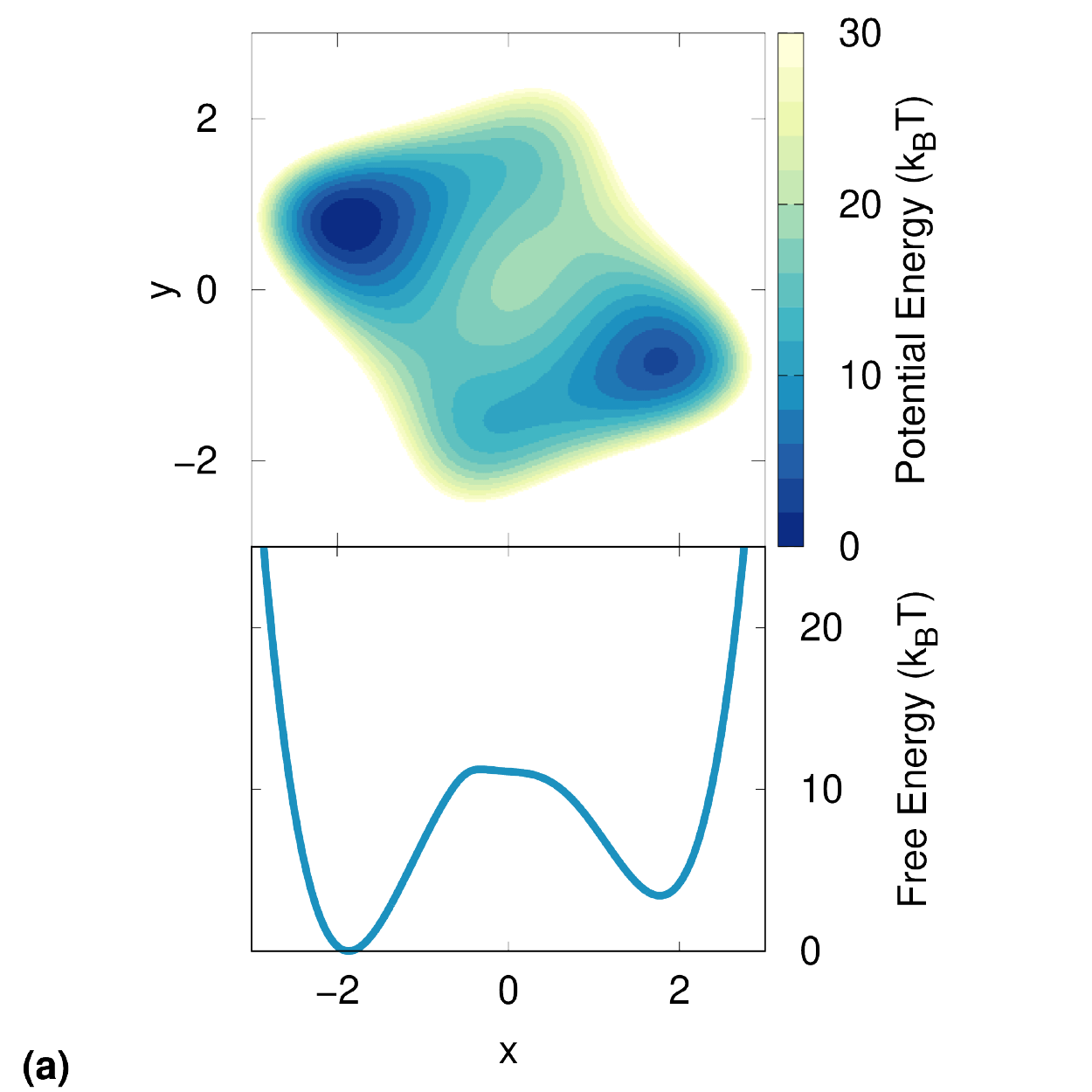}
        & \includegraphics[width=0.45\textwidth]{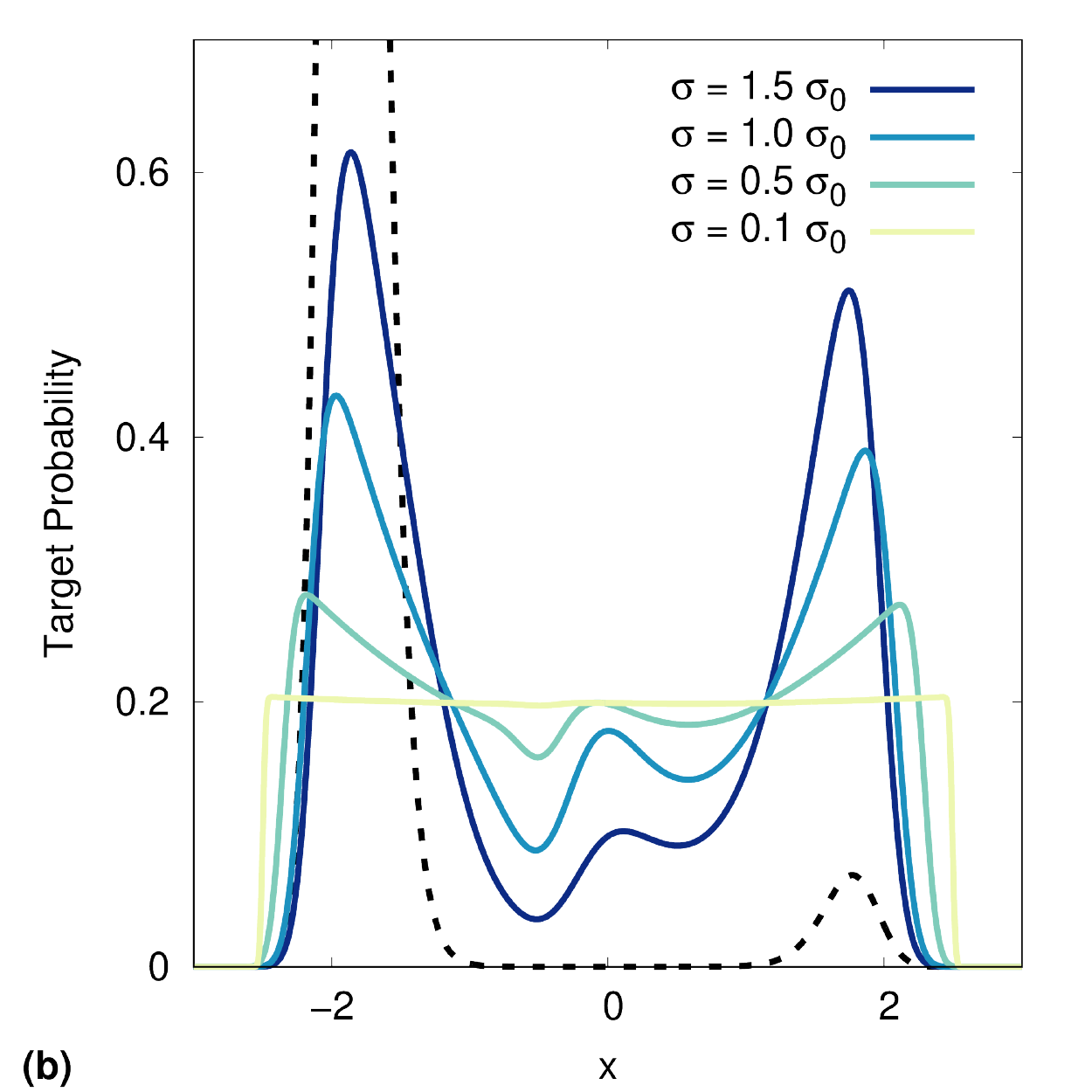}
    \end{tabular}
    \caption{Double-well model in the multiumbrella ensemble. 
    (a) Potential energy of the double-well 2D model system, and its free energy along the $x$ coordinate.
    (b) Multiumbrella target distribution, for different values of umbrella width $\sigma$.
    The black dotted line is the unbiased probability distribution $P_0(x)$.
  \label{F:model}}
\end{figure*}

As an example for the multiumbrella ensemble we consider a Langevin dynamics on a 2D model potential\cite{Invernizzi2019} using as CV the $x$ coordinate only, Fig.~\ref{F:model}a.
Such CV is suboptimal, in the sense that it does not include all the slow modes of the system, and can be problematic when performing standard umbrella sampling with multiple windows.
For instance the window centered at $x=0$ cannot be efficiently sampled in a single simulation, since it presents a barrier along $y$.
With our approach this suboptimality only causes a slower convergence, but does not constitute a problem, and no extra care is required to handle it.

Figure \ref{F:model}b shows how the target distribution changes for different $\sigma$ choices, expressed in units of the unbiased standard deviation in the basins, $\sigma_0\approx 0.18$ .
The FES estimate could be directly obtained from the $\Delta F_n(s)$, but in the case of large $\sigma$ this would lead to an estimate in which features are oversmoothed (see \href{https://arxiv.org/src/2007.03055v3/anc/SupplementalMaterial.pdf}{SM}).
As a general rule it is better to estimate the FES via the reweighting procedure.

In the \href{https://arxiv.org/src/2007.03055v3/anc/SupplementalMaterial.pdf}{supplemental material} we provide all the simulation details and show the convergence of the free energy, comparing it to well-tempered OPES and metadynamics.
While in metadynamics and well-tempered OPES the bias is constructed in such a way to push the system out of the visited areas, with multiumbrella OPES we are forcing the system to stay in a chosen CV region.
Despite this difference in both cases we have similar target distributions and the resulting sampling allows us to reconstruct the FES.

\subsection{Combining Thermodynamic and\\Order Parameter Expansions\label{S:combining}}
An important characteristic of the present scheme is that it allows for a straightforward combination of different expanded ensembles.
In particular it allows for a rigorous and efficient combination of thermodynamic generalized ensembles with enhanced sampling along a system-specific order parameter.

To understand why this is important one can think about a first order phase transition, where there is a kinetic bottleneck between the two phases that is responsible of an ergodicity problem.
Increasing the temperature typically changes the relative stability of the two phases, but the free energy barrier separating them might remain high along the whole coexistence line, thus making convergence very slow.
A possible solution is to identify a suitable order parameter and biasing it to increase the transition probability.
Combining the two approaches might actually outperform both\cite{Yang2018,Niu2019}.
This kind of combination can be useful not only for phase transitions, for instance also in alchemical free energy calculations an open problem is how to properly handle barriers orthogonal to the transformation\cite{Hahn2020}.

We have already cited some hybrid methods that combine a replica-exchange approach with metadynamics, in order to enhance the sampling both along a thermodynamic quantity and an order parameter\cite{Bussi2006,Piana2007,Bonomi2010,Nava2015,Yang2018}.
A non-hybrid approach has been first proposed with multidimensional replica exchange\cite{Sugita2000}, but it has the drawback of requiring a sometimes impractical number of parallel replicas, due to the multidimensionality of the expansion.
With OPES we can sample the same target distribution of multidimensional replica exchange, but using a bias potential and without requiring a minimum number of parallel replicas.
In developing our method we followed the footsteps of another non-hybrid approach that has been recently proposed by our group, using the VES formalism and a custom target distribution\cite{Piaggi2019cv,Niu2020}.
Compared to the very flexible and customizable VES approach, OPES has the advantage of having much fewer free parameters and thus being simpler to set up and use.

\subsubsection*{Example: Sodium}
We consider here as an example the calculation of the liquid-bcc phase diagram of a model of sodium\cite{Mendelev2015}, the same studied in Ref.~\citenum{Piaggi2019cv}.
We will sample the liquid and solid phase over a range of temperatures and pressures, using a recently proposed order parameter $s$, called environment similarity collective variable\cite{Piaggi2019cv}.
Such CV provides a measure of the crystallinity of the system, by comparing the local environment of the atoms to a reference one.
For this reason we will refer to it as crystallinity CV, but it is actually more general and can be used to describe a variety of phase transitions\cite{Niu2020,Piaggi2020}.

Using LAMMPS patched with PLUMED, we perform NPT simulations, $u_0(\mathbf{x})=\beta_0 U(\mathbf{x})+\beta_0 p_0 V(\mathbf{x})$.
We can write the OPES equations, Eqs.~(\ref{E:bias_n}) and Eqs.~(\ref{E:deltaF_n}), via the following expansion CVs
\begin{equation}
    \Delta u_{\beta,p,s}(\mathbf{x})=(\beta-\beta_0)U(\mathbf{x})+(\beta p-\beta_0 p_0)V+\frac{(s(\mathbf{x})-s)^2}{2\sigma^2}\, .
\end{equation}
The free energy estimates $\Delta F_n(\beta,p,s)$ are expressed as a function of the inverse temperature $\beta$, the pressure $p$ and the crystallinity CV $s$.
The bias $v=v(U,V,s)$, is expressed as a function of the potential energy $U$, the volume $V$ and the crystallinity CV $s$.

The simulation is performed with 250 atoms at $T_0=400$~K and $p_0=0.5$~GPa (5 kbar), using 4 multiple walkers that share the same bias and contribute to the same ensemble averages to update the $\Delta F_n(\beta,p,s)$ estimate.
The aim is to sample liquid and solid configurations in the temperature range from 350~K to 450~K and pressures from 0~GPa to 1~GPa (10 kbar).
The uniform grid over $\beta$ and $p$ to define the target distribution is automatically generated from a short 100~ps unbiased run, and consists of 4 temperature steps and 8 pressure steps.
We chose as $\sigma$ for the multiumbrella target a value of about 2.5 times the unbiased standard deviation in the basins, and it determines the presence of 26 umbrellas uniformly placed between $s_{\min}=0$ (liquid) and $s_{\max}=1$ (solid).
In total the $\Delta F_n(\beta,p,s)$ to be estimated are $4\times8\times26=832$.
After less than 3~ns the bias is practically converged, but the simulation is kept running until a total combined time of 100~ns, in order to collect enough statistic for a smooth reweighting.

\begin{figure*}
   \begin{tabular}{cc}
  \includegraphics[width=0.45\textwidth]{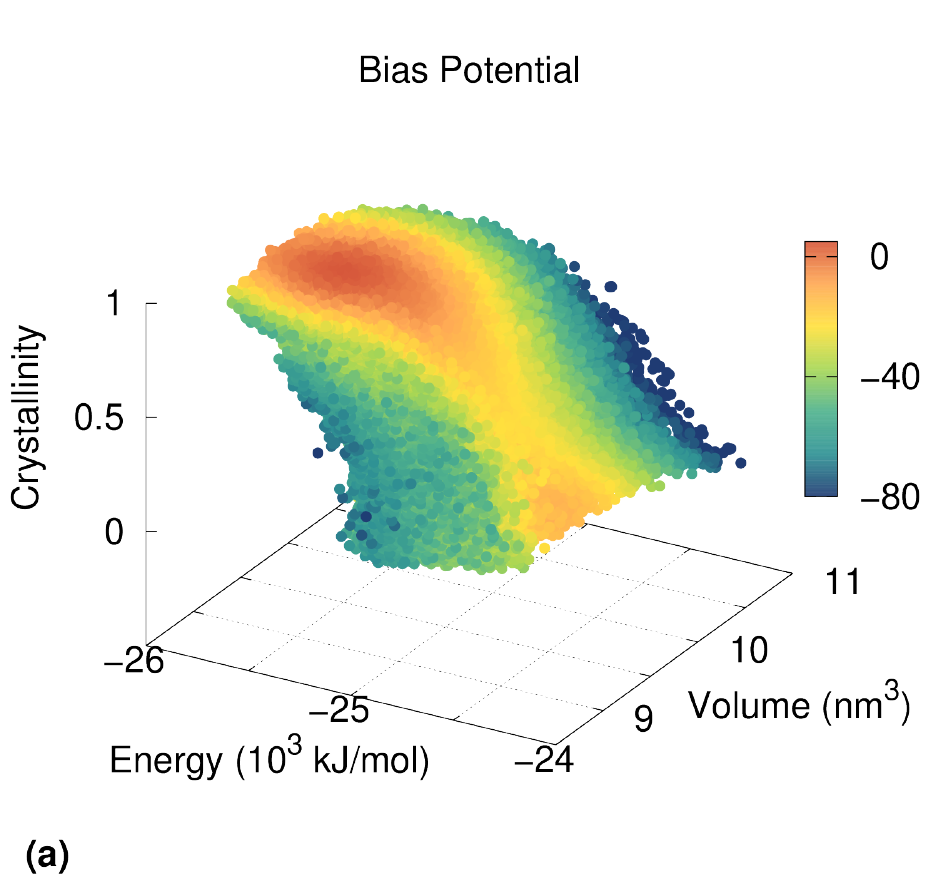}
  & \includegraphics[width=0.45\textwidth]{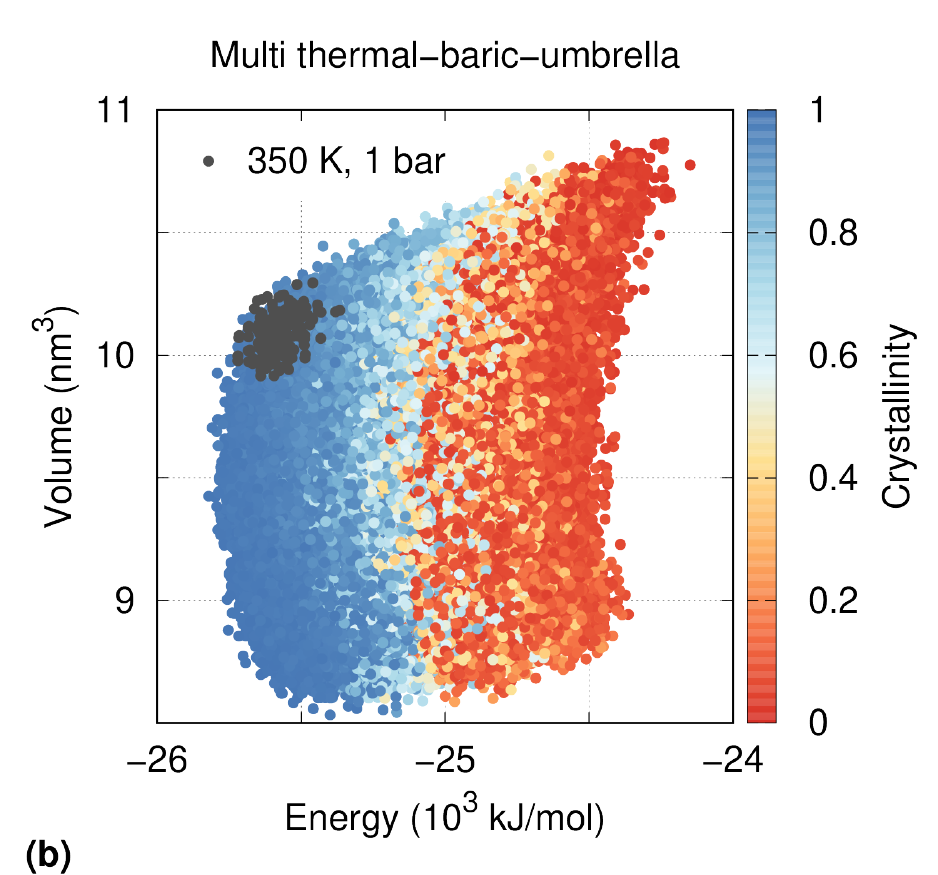}
  \end{tabular}
  \caption{Configurations of sodium sampled during the multithermal-multibaric-multiumbrella simulation.
  (a) The points are colored accordingly to the value of the bias $v(U,V,s)$.
  (b) The points are shown in the energy-volume space and colored accordingly to the value of the crystallinity CV $s$.
  As reference we also show in grey the region sampled during an unbiased simulation in the bcc phase.
  \label{F:sodium-colvar}}
\end{figure*}
In Fig.~\ref{F:sodium-colvar}a we show the points sampled in the CV space during the simulation, colored according to the value of the bias potential $v(U,V,s)$.
We can clearly distinguish the hourglass shape described in Ref.~\citenum{Piaggi2019cv}.
In Fig.~\ref{F:sodium-colvar}b we see the same trajectory plotted on the energy-volume plane.
For comparison we show the configuration sampled in an unbiased simulation at a single temperature and pressure, in which the system remains all the time in the bcc crystal phase.

\begin{figure*}
   \begin{tabular}{cc}
  \includegraphics[width=0.45\textwidth]{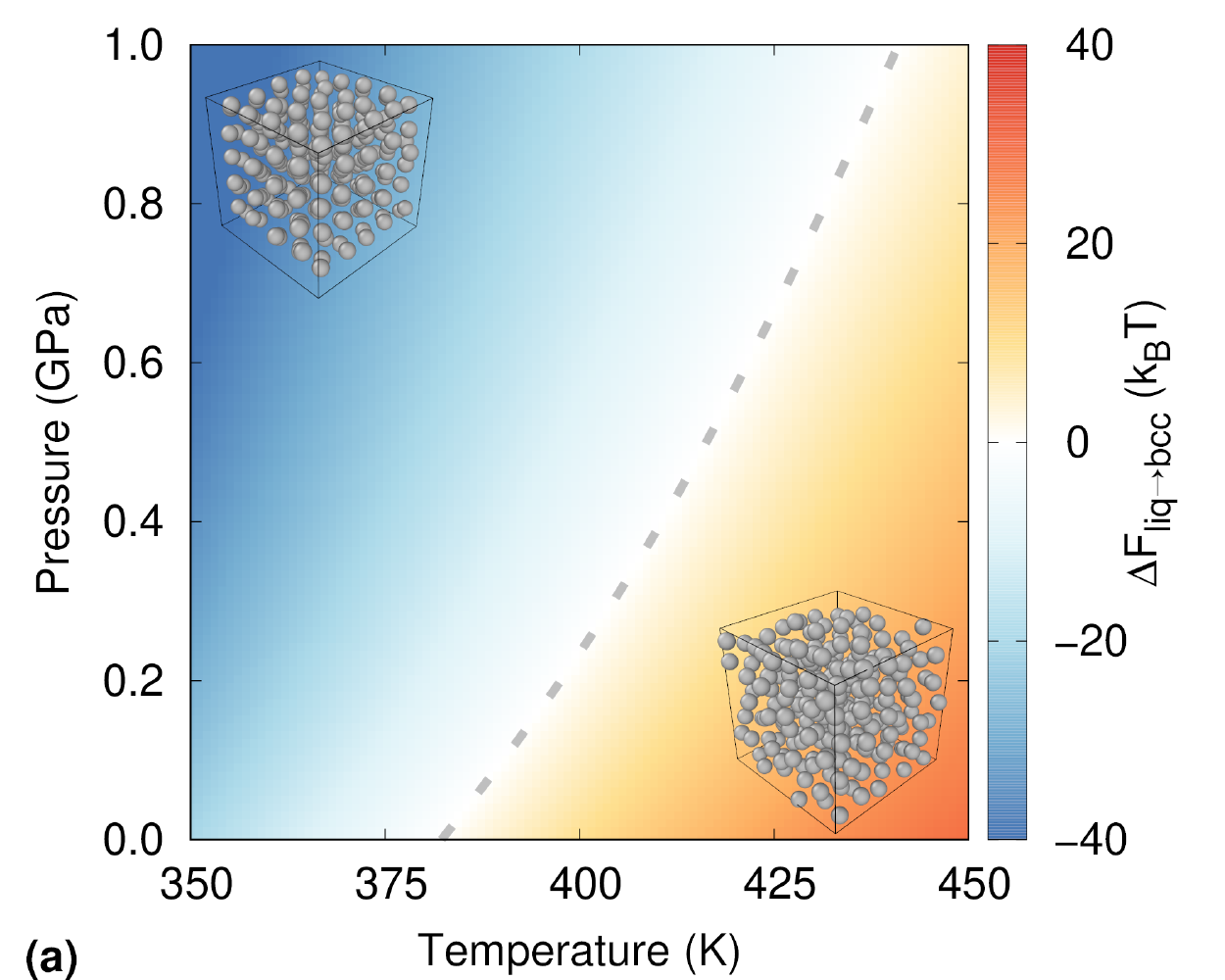}
  & \includegraphics[width=0.45\textwidth]{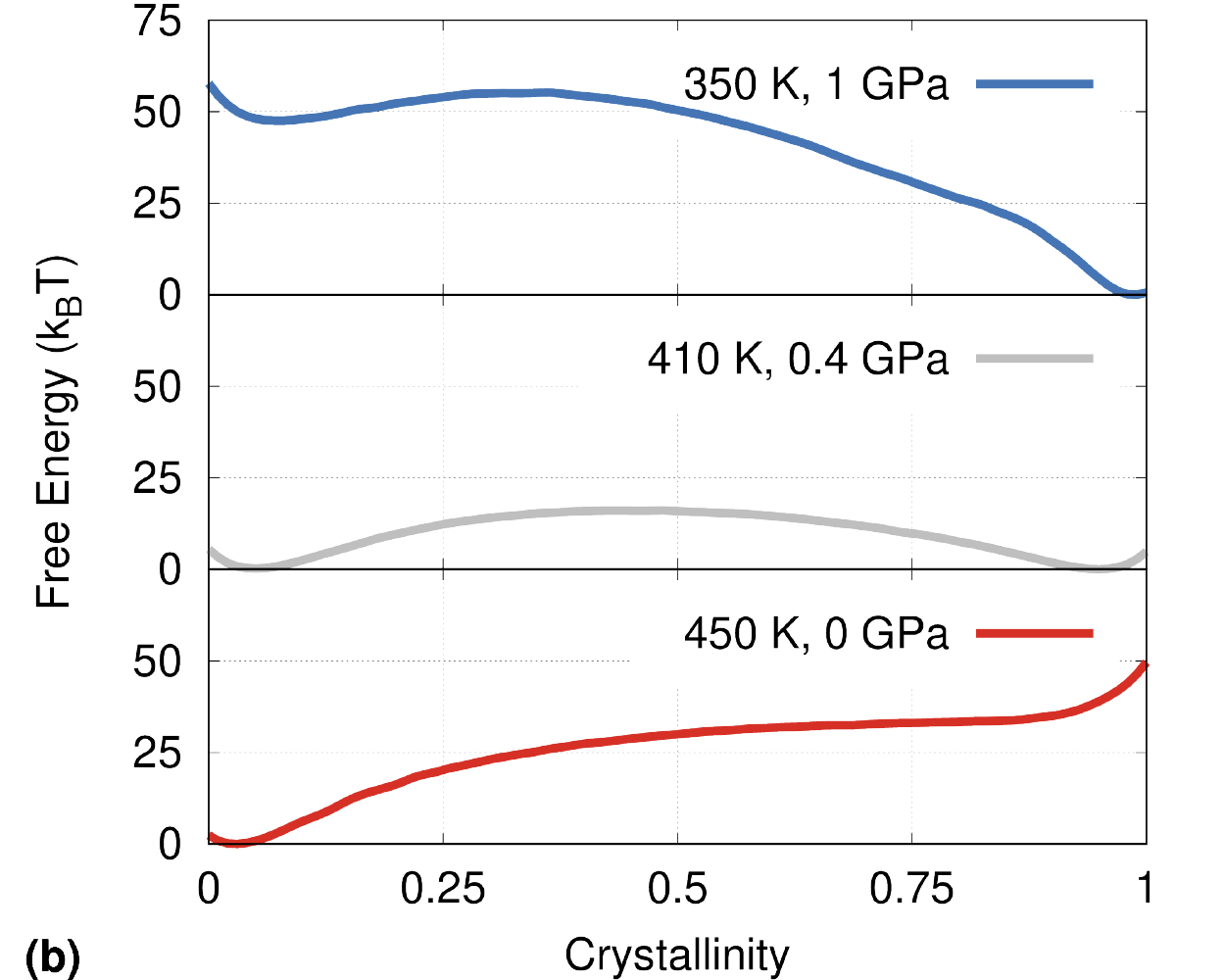}
  \end{tabular}
  \caption{Phase equilibrium of liquid and solid (bcc) sodium using a combination of thermodynamic and order parameter expansions.
  (a) Free energy difference between the phases, $\Delta F_{liq \rightarrow bcc}$, at different thermodynamic conditions.
  The coexistence line is shown as a grey dashed line.
  (b) Free energy surfaces as a function of the crystallinity CV, for three representative thermodynamic conditions.
  Error bars are smaller than the line width.
  \label{F:sodium}}
\end{figure*}

We can define the free energy difference between the two phases as in Ref.~\citenum{Piaggi2019cv}:
\begin{equation}
  \Delta F_{liq \rightarrow bcc}(T,p)=-\log \left(\frac{\langle \chi_{s\in[0.5,1]} \rangle_{T,p}}{\langle \chi_{s\in[0,0.5]} \rangle_{T,p}}\right)\, ,
\end{equation}
where $\chi$ is a characteristic function, equal to 1 if the variable is in the proper range and 0 otherwise, and $\langle \cdot \rangle_{T,p}$ is the ensemble average at temperature $T$ and pressure $p$.

In Fig.~\ref{F:sodium}a we show $\Delta F_{liq \rightarrow bcc}(T,p)$ obtained by reweighting at different temperatures and pressures. 
The coexistence line $\Delta F_{liq \rightarrow bcc}(T,p)=0$ is shown with a dotted gray line.
On the right side, Fig.~\ref{F:sodium}b, we provide the free energy surface as a function of the CV, $F(s)$, at different thermodynamic conditions.
Error bars are calculated with a weighted block average (Appendix~\ref{A:block}) and all the results are in agreement with Ref.~\citenum{Piaggi2019cv}, see \href{https://arxiv.org/src/2007.03055v3/anc/SupplementalMaterial.pdf}{SM}.

It is important to notice that while the relative stability between liquid and solid changes across the range considered here, the probability of being in the transition state between the two is always extremely small, as can be seen from the high FES values around $s=0.5$ in Fig.~\ref{F:sodium}b.
By actively biasing the CV $s$ we allow the system to efficiently sample the transition region as well, and this makes it possible to quickly converge the multithermal-multibaric simulation despite the presence of a first order phase transition.

\section{About the Optimal\\Target Distribution\label{S:optimal}}
Before reaching the conclusion of the paper we would like to add a final remark.
At the beginning of Sec.~\ref{S:targeting} we presented the non-weighted expanded ensemble target distribution, $p_{\{\lambda\}}(\mathbf{x})=\frac{1}{N_{\{\lambda\}}}\sum_\lambda P_\lambda(\mathbf{x})$.
It is reasonable to wonder if this is an optimal target and in which sense.
We argue here that the effective sample size can be used to define an optimality criterion.

Let's say our goal is to sample from a generalized ensemble that contains all the relevant microscopical configurations for a give range of the parameter $\lambda$.
While the expanded target distribution $p_{\{\lambda\}}(\mathbf{x})$, Eq.~(\ref{E:target}), fulfills such goal, there are in principle other possible choices.

It is useful to look at the special case of multicanonical ensembles, that has a long history (see also Sec.~\ref{S:multicanonical}).
In this context various different target distributions have been used, other than the non-weighted expanded ensembles one.
One option is to have a uniform sampling in the energy\cite{Berg1991,Lee1993,Piaggi2019}, another one is to have a uniform sampling in the entropy\cite{Hesselbo1995}, and a third one is to define the target by integrating the probability over the temperature, as in Refs.~\citenum{Gao2008,Martinsson2019}.
In this last approach one often approximates the integral with a sum, and effectively uses a target similar to our, Eq.~(\ref{E:target}), which is also the one used in temperature replica exchange.
Another interesting perspective is presented in Ref.~\citenum{Lindahl2018}, where a Riemann metric is introduced to define optimality.

We believe that if our goal is to reweight at different temperatures, then the optimal target distribution is the one that yields the highest possible uniform effective sample size over the whole considered $\Delta \lambda$ range.
Here we will not further explore such optimal target distribution nor dig deeper in the definition of effective sample size.
However, simply by defining this criterion we can notice that one should not see the sum in Eq.~(\ref{E:target}) as an approximation to an integral.
As a matter of fact, using as few as possible intermediate $\lambda$-points brings us closer to this optimal target than having more, at least in the systems we studied (see \href{https://arxiv.org/src/2007.03055v3/anc/SupplementalMaterial.pdf}{SM}).

It might also be the case that one is not interested in obtaining statistic for the whole $\Delta \lambda$ range, but only for a subset of $\lambda$-states.
In this case the optimal target would be the one that maximizes the effective sample size for those $\lambda$-states while ensuring ergodicity.
According to this criterion we argued in Ref.~\citenum{Invernizzi2020} that the well-tempered target is better than a uniform one, since it allows for an ergodic sampling while providing a higher $n_{\text{eff}}/n$ ratio and avoids unimportant high free energy regions.

\section{Conclusion}
In this paper we presented a general framework that provides a unified approach to enhanced sampling.
To implement our method we leveraged the iterative scheme of OPES, an enhanced sampling method based on the construction of a bias potential along a set of collective variables, that was originally introduced for metadynamics-like sampling.
We showed how this approach can be used to sample the same expanded ensembles typically sampled by a different family of enhanced sampling methods.

We also introduced the concept of expansion CVs, $\Delta u_\lambda(\mathbf{x})$, that can be used to fully characterize a non-weighted expanded target distribution $p_{\{\lambda\}}(\mathbf{x})$, Eq.~(\ref{E:target}), together with the free energy differences to be iteratively estimated, Eq.~(\ref{E:deltaF_def}), and the target bias, Eq.~(\ref{E:bias_def}).

We then presented various examples of the application of the method to sample the most common expanded ensembles.
These ensembles are summarized in Tab.~\ref{T:ensembles}.
In particular we have shown how OPES can be used to enhance at the same time temperature-related fluctuations and a system-specific order parameter, Sec.~\ref{S:combining}.

We notice that in defining the target distribution $p^{tg}(\mathbf{x})$ we consider only the positional degrees of freedom, and not the atomic velocities.
Thus the ensembles sampled by our method are not identical to the ones sampled for instance by replica exchange, even though the target distribution is the same.
In fact the two methods sample the same configuration space, but a different velocity space.
This does not have an effect on any statistical average of observables that are function of the coordinates only, but might be an interesting point for further research.

In the future it would be interesting to combine expanded target distributions with well-tempered-like distributions, which can scale better with higher dimensionality.
Also weighted expanded targets might be of interest, where each sub-ensemble $\lambda$ has a specific different normalized weight.
More generally, we believe that our perspective of focusing on the target distribution has further potential that should be explored.

\begin{table*}
  \caption{Some of the most common expanded ensembles, together with the expansion collective variables $\Delta u_\lambda(\mathbf{x})$ that define the OPES target bias, Eq.~(\ref{E:bias_def}), and the free energy differences $\Delta F(\lambda)$, Eq.~(\ref{E:deltaF_def}).
  Each of the considered target biases can in turn be expressed as a function of one or two CVs.
  It is also possible to easily combine these ensembles to form new ones, as shown in Sec.~\ref{S:combining}.
  \label{T:ensembles}}
\begin{ruledtabular}
\begin{tabular}{lccc}
     Target Ensemble & Expansion CVs & Parameters & CVs \\
     \hline
     Linearly Expanded & $\lambda \Delta u(\mathbf{x})$ & $\{\lambda\}$ & $\Delta u(\mathbf{x})$ \\
     Multicanonical & $(\beta-\beta_0)U(\mathbf{x})$ & $\{\beta\}$ & $U(\mathbf{x})$ \\
     Multibaric & $\beta_0 (p-p_0) V(\mathbf{x})$ & $\{p\}$ & $V(\mathbf{x})$ \\
     Multithermal-Multibaric & $(\beta-\beta_0)U(\mathbf{x}) +(\beta p-\beta_0 p_0) V(\mathbf{x})$ & $\{\beta, p\}$ & $U(\mathbf{x}), V(\mathbf{x})$ \\
     Multiumbrella & $(s(\mathbf{x})-s_\lambda)^2/(2\sigma^2)$ & $\{s_\lambda\}$ & $s(\mathbf{x})$\\
\end{tabular}
\end{ruledtabular}
\end{table*}

\section*{Data Availability Statement}
The method is implemented in the open source PLUMED enhanced sampling library\cite{plumed}, and will be available as a contributed module called OPES.
All the input files and post-processing scripts used for this paper are openly available on the PLUMED-NEST\cite{nest} (\url{www.plumed-nest.org} plumID:20.022), and in the Materials Cloud Archive (\url{www.materialscloud.org} materialscloud:2020.81), where also the trajectories of the simulations are stored.

\begin{acknowledgments}
The authors thank Riccardo Capelli and Sandro Bottaro for useful discussions, and Andrea Rizzi for helpful feedback on the manuscript.

This research was supported by the NCCR MARVEL, funded by the Swiss National Science Foundation, and European Union Grant No.~ERC-2014-AdG-670227/VARMET.
P.M.P. acknowledges support from the Swiss National Science Foundation through an Early Postdoc.Mobility fellowship.
Calculations were carried out on Euler cluster at ETH Zurich and in Swiss National Supercomputing Center (CSCS) cluster Piz Daint.

\end{acknowledgments}

\appendix

\section{Weighted Block Average\label{A:block}}
It is well know that when performing molecular dynamics one should take into account for the time correlation of the samples in order to compute the uncertainty of a given estimate.
Methods such as block averaging\cite{Flyvbjerg1989} are commonly used to properly handle this.
The effect of such time correlation is to make the sample size effectively smaller, thus simply taking the square root of the variance divided by the number of samples would underestimate the actual uncertainty.
When dealing with weighted samples, as it is the case when a bias potential is used, the effective sample size is further reduced by the presence of these weights.
We must take into account also for this effect when we perform block averaging.
We report here the protocol that we follow to estimate uncertainties, which is the same as the one presented in Ref~\citenum{Bussi2019}, but here we highlight the role played by the effective sample size.

We are interested in estimating the ensemble average of an observable $O=O(\mathbf{x})$ from a biased ensemble
\begin{equation}
  \langle O \rangle \approx \hat{O}=\frac{\sum_{k=1}^n O_k w_k}{\sum_{k=1}^n w_k}\, ,
\end{equation}
where $w_k$ are the weights due to the bias potential, as in Eq.~(\ref{E:reweight}).
In order to estimate the uncertainty we divide the data into $M$ sub-sets or blocks, each containing an equal number of samples $n/M$.
We then calculate the estimate $\hat{O}_i$ from each block, via a weighted average, and also the weight of the $i$th block
\begin{equation}
    W_i=\sum_{k=(i-1) (n/M)}^{i (n/M)} w_k \, .
\end{equation}
In this way the total estimate can be obtained as the weighted average of the blocks,
\begin{equation}
    \hat{O}=\frac{\sum_{i=1}^M W_i \hat{O}_i}{\sum_{i=1}^M W_i}\, .
\end{equation}
Then according to the usual block average procedure we estimate the unbiased variance between the blocks, that in this case is a weighted variance:
\begin{equation}
    \sigma^2_O=\frac{M_{\text{eff}}}{M_{\text{eff}}-1}\frac{\sum_{i=1}^M W_i (\hat{O}_i-\hat{O})^2}{\sum_{i=1}^M W_i}\, ,
\end{equation}
where instead of the total number of blocks $M$, we use the effective block size $M_{\text{eff}}<M$:
\begin{equation}
    M_{\text{eff}}=\frac{\left(\sum_{i=1}^M W_i\right)^2}{\sum_{i=1}^M W_i^2}\, ,
\end{equation}
that is the same as Eq.~(\ref{E:neff}).
The statistical error on the $\hat{O}$ estimate is then $\sigma_O/\sqrt{M_{\text{eff}}}$. 
When the number of blocks $M$ is small, or when the weight $W_i$ are unbalanced, using $M$ instead of $M_{\text{eff}}$ can introduce a considerable underestimate of the real uncertainty.

The usual block average procedure can then be carried out, repeating the analysis using different number of blocks $M$ and looking for a plateau in the error estimate.

\section{Improving Robustness for\\the Multiumbrella Target\label{A:robustness}}
In order to make the iterative optimization scheme more effective, in some cases it proved useful to introduce two small modifications when targeting the multiumbrella ensemble.

When the iterative scheme starts, the first guess of the $\Delta F_n(s_\lambda)$ comes from just one single point, and is thus very inaccurate for CV values far away from the visited one.
In particular it tends to become extremely large, which causes the bias to be very strong in pushing the system to the farthest $s_\lambda$-value.
This initial bias is stronger the smaller the $\sigma$, and might even cause the simulation to fail during the very first biased steps.
To avoid this, we can limit the initial value of the $\Delta F_n(s_\lambda)$ estimates to be always smaller than a given value, thus $\Delta F_0(s_\lambda) \le \Delta E$.
This $\Delta E$ value can be set to be equal to an estimate of the free energy barrier that has to be overcome.
Thus, similarly to Ref.~\citenum{Invernizzi2020}, we add an extra optional parameter called ``barrier'', that sets the value of $\Delta E$.
This barrier guess does not have to be perfect and a very rough estimate typically suffices.
Also, we did not observe significant change in the convergence speed, thus we suggest to use this extra parameter only in case of an initial failure of the simulation, due to a too strong initial bias.
The barrier parameter can in principle be used also with types of expansion other than the multiumbrella one, even though in those cases it might be less useful.

The second modification, comes from the observation that, contrary to the previously considered expanded ensembles, the multiumbrella one is not granted to sample the full unbiased distribution $P_0(\mathbf{x})$.
This can be a problem for the iterative scheme, because all the $\Delta F_n(s_\lambda)$ use as reference the unbiased free  energy $F_0=-\log Z_0$, whose estimate can vary substantially if $P_0$ is not properly sampled.
Typically it should be easy to chose an interval $\Delta s$ that contains all the $s$ values relevant for $P_0$, but if this is not the case then the problem can be fixed by simply adding $P_0$ itself to the target distribution.
To add $P_0$ to the target, it is sufficient to add an extra expansion CV $\Delta u_0(\mathbf{x}) \equiv 0$, that always returns zero.

In the examples presented in the paper we did not have to use any of these two modifications.

\bibliography{biblio}

\end{document}